\documentclass[twocolumn]{aastex63}
\usepackage{natbib}
\usepackage{apjfonts}
\usepackage{graphicx}
\usepackage{epstopdf}

\newcommand{\chandra}{{Chandra\/}}
\newcommand{\xmm}{\hbox{XMM-Newton\/}}
\newcommand{\nustar}{{NuSTAR\/}}
\newcommand{\flux}{{erg~cm$^{-2}$~s$^{-1}$}}

\newcommand\iona[2]{#1$\;${\scshape{#2}}}
\newcommand{\nh}{$N_{\mbox{\scriptsize H}}$}
\newcommand{\aspc}{ASP Conf. Ser.}
\newcommand{\xray}{{\hbox{X-ray}}}
\usepackage{bm}
\begin{document}

\title{NuSTAR Observations of Intrinsically 
\hbox{X-ray} Weak Quasar Candidates: An Obscuration-Only Scenario}

\author{Chaojun~Wang}
\affiliation{School of Astronomy and Space Science, Nanjing University, Nanjing, Jiangsu 210093, China}
\affiliation{Key Laboratory of Modern Astronomy and Astrophysics (Nanjing University), Ministry of Education, Nanjing 210093, China}

\author{B.~Luo}
\affiliation{School of Astronomy and Space Science, Nanjing University, Nanjing, Jiangsu 210093, China}
\affiliation{Key Laboratory of Modern Astronomy and Astrophysics (Nanjing University), Ministry of Education, Nanjing 210093, China}

\author{W.~N.~Brandt}
\affiliation{Department of Astronomy \& Astrophysics, 525 Davey Lab,
The Pennsylvania State University, University Park, PA 16802, USA}
\affiliation{Institute for Gravitation and the Cosmos, The Pennsylvania State University,
University Park, PA 16802, USA}
\affiliation{Department of Physics, 104 Davey Lab, The Pennsylvania State University, University Park, PA 16802, USA}

\author{D.~M.~Alexander}
\affiliation{Centre for Extragalactic Astronomy, Department of Physics, Durham University, Durham DH1 3LE, UK}

\author{F.~E.~Bauer}
\affiliation{Instituto de Astrof{\'{\i}}sica and Centro de Astroingenier{\'{\i}}a, Facultad de F{\'{i}}sica, Pontificia Universidad Cat{\'{o}}lica de Chile, Casilla 306, Santiago 22, Chile}
\affiliation{Millennium Institute of Astrophysics, Nuncio Monse{\~{n}}or S{\'{o}}tero Sanz 100, Of 104, Providencia, Santiago, Chile}
\affiliation{Space Science Institute, 4750 Walnut Street, Suite 205, Boulder, Colorado 80301, USA}

\author{S.~C.~Gallagher}
\affiliation{Department of Physics \& Astronomy and Institute for Earth and Space Exploration, The University of Western
  Ontario, London, ON, N6A 3K7, Canada}

\author{Jian Huang}
\affiliation{School of Astronomy and Space Science, Nanjing University, Nanjing, Jiangsu 210093, China}
\affiliation{Key Laboratory of Modern Astronomy and Astrophysics (Nanjing University), Ministry of Education, Nanjing 210093, China}

\author{Hezhen Liu}
\affiliation{School of Astronomy and Space Science, Nanjing University, Nanjing, Jiangsu 210093, China}
\affiliation{Key Laboratory of Modern Astronomy and Astrophysics (Nanjing University), Ministry of Education, Nanjing 210093, China}

\author{D.~Stern}
\affiliation{Jet Propulsion Laboratory, California Institute of Technology, 4800 Oak Grove Drive, MS 169-224, Pasadena, CA 91109, USA}

\begin{abstract}
We utilize recent NuSTAR observations (co-added depth $\approx55$--120~ks) 
of PG $1001+054$, PG $1254+047$, and PHL 1811 to constrain their hard X-ray ($\gtrsim5$~keV) weakness and spectral
shapes, and thus
to investigate the nature of their extreme X-ray weakness. 
These quasars showed very weak soft X-ray emission, and they
were proposed to be intrinsically X-ray weak, with the X-ray coronae
producing weak continuum emission relative to their optical/UV emission.
However, the new observations suggest an alternative explanation.
The \nustar\ 3--24~keV spectral shapes for PG $1001+054$ and
PHL~1811 are likely flat (effective power-law photon indices 
$\Gamma_{\rm eff}=1.0^{+0.5}_{-0.6}$ and
$\Gamma_{\rm eff}=1.4^{+0.8}_{-0.7}$, respectively), while the shape is nominal
for PG~$1254+047$ ($\Gamma_{\rm eff}=1.8\pm0.3$).
PG~$1001+054$ and PHL~1811 are significantly weak at hard X-ray energies
(by factors of $\approx26$--74 at
rest-frame 8~keV)
compared
to the expectations from their optical/UV emission, while
PG $1254+047$ is only hard X-ray weak by a factor of $\approx3$. 
We suggest that X-ray obscuration is present in all
three quasars.
We propose that, as an
alternative to the intrinsic 
X-ray weakness $+$ \hbox{X-ray} obscuration scenario, 
the soft and hard X-ray weakness of these quasars can 
be uniformly explained under an obscuration-only scenario.
This model provides adequate 
descriptions of the multi-epoch soft and hard \xray\ data of
these quasars, with variable column density and leaked fraction of the partial-covering absorber.
We suggest that the absorber is the clumpy dust-free wind launched from 
the accretion disk. 
These quasars probably have super-Eddington accretion 
rates that drive powerful and high-density winds.

\end{abstract}


\section{INTRODUCTION}

Active galactic nuclei (AGNs) generally produce luminous X-ray emission, 
which is believed to originate largely from the accretion-disk corona in 
the vicinity of the central supermassive black hole (SMBH) 
via inverse-Compton scattering of the optical/UV seed photons from the accretion disk 
(e.g., \citealt{Turner2009,Done2010,Gilfanov2014,Fabian2017}). 
AGN X-ray continua typically have an intrinsic 
power-law shape ($N_E \propto E^{-\Gamma}$),
and the mean value of the photon indices ($\Gamma$) for radio-quiet AGNs is around 
\hbox{1.9--2.0} with a scatter of $\approx 0.2$ \cite[e.g.,][]{Reeves1997,Just2007,Scott2011}.
Observations of the X-ray and UV emission from large samples of radio-quiet AGNs 
have revealed that the \hbox{X-ray} flux is closely correlated
with the optical/UV flux, indicating a strong physical connection between the accretion 
disk and the X-ray corona.
The correlation is typically expressed as a negative relation
between the X-ray-to-optical
\hbox{power-law} slope parameter ($\alpha_{\rm OX}$)\footnote{$\alpha_{\rm OX}$ is
defined as $\alpha_{\rm OX}$ = 0.3838log($f_{\rm 2keV}$/$f_{\rm 2500~{\textup{\AA}}}$) 
\citep{Tananbaum1979}, where $f_{\rm 2keV}$ and $f_{\rm 2500~{\textup{\AA}}}$ are 
the \hbox{rest-frame} 2~keV and 
2500~\AA\ flux densities.}
and the ${2500~{\textup{\AA}}}$
monochromatic luminosity ($L_{\rm 2500~{\textup{\AA}}}$), and it is highly significant
across a broad population of AGNs,
ranging from moderate-luminosity 
AGNs to the most luminous quasars
(e.g., \citealt{Strateva2005,steffen2006,Just2007,Lusso2017,Liu2021}).

The observed X-ray emission from AGNs may be modified by line-of-sight obscuration,
resulting in lower observed \hbox{X-ray} fluxes than those expected from the 
$\alpha_{\rm OX}\textrm{--}L_{2500~\textup{\AA}}$ relation. 
A common approach to parameterize the amount of 
\xray\ weakness uses
the $\Delta\alpha_{\rm OX}$ parameter,
defined as the difference between
the observed
$\alpha_{\rm OX}$ value and the $\alpha_{\rm OX}$ value expected from
the $\alpha_{\rm OX}$--$L_{\rm 2500~{\textup{\AA}}}$ relation; 
$\Delta\alpha_{\rm OX}=-0.3838$ thus corresponds to a 
factor of X-ray weakness of 10 at rest-frame 2 keV.
Type 2 AGNs are
generally \xray\ obscured, likely due to the dusty ``torus''
\citep[e.g.,][]{Netzer2015,Hickox2018}.
Type 1 AGNs may also have X-ray obscuration from largely dust-free gas.\footnote{Similar 
obscuration from dust-free gas might also be present in some of the type 2 AGNs, though
usually not distinguishable from the torus obscuration.} 
For example,
broad absorption line (BAL) quasars, which are characterized by blueshifted
broad UV absorption lines (e.g., \iona{C}{iv} $\lambda1549$), generally 
show weak X-ray emission \cite[e.g.,][]{Gallagher2002,Gallagher2006,Fan2009,Gibson2009}.
One frequently adopted physical model for BAL quasars is the \hbox{disk-wind} model,
where the observed BALs originate from an outflowing 
equatorial wind launched from the accretion disk
and radiatively driven by UV-line pressure \citep[e.g.,][]{Murray1995,Proga2000,Matthews2016}.
This model usually invokes
``shielding'' gas between the wind and nucleus
or a clumpy wind \citep[e.g.,][]{Baskin2014,Matthews2016,Giustini2019}
to provide obscuration of the nuclear extreme UV (EUV) and \xray\ radiation
which might otherwise overionize the wind and hamper radiative
acceleration. BAL quasars are considered generally to have larger inclination angles
than non-BAL quasars, with the line of sight to the UV continuum region of the accretion disk
intersecting the wind, leading to the observed BALs. The line of sight to the
X-ray emitting corona, though not necessarily the same as the UV line of sight, is likely 
also through the shielding gas or the clumpy wind, resulting
in the often observed X-ray weakness (e.g., Figure~1 of \citealt{Luo2013}). 
Besides BAL quasars, a small fraction ($5.8\pm0.7\%$) of 
non-BAL type~1 quasars have been found to be X-ray weak, likely due to absorption 
\citep[e.g.,][]{Pu2020}.
They may share a similar nature to the BAL quasars; they do not show any UV BALs 
probably due to geometric effects (e.g., small inclination angles) or 
a low velocity of the wind along the UV line of sight \citep[e.g.,][]{Giustini2019}.

A small subset of AGNs has been proposed to be intrinsically \hbox{X-ray} weak, 
producing much less
X-ray emission than expected from
the \hbox{$\alpha_{\rm OX}$--$L_{\rm 2500{\textup{\AA}}}$}
relation. 
These candidates are observed to be significantly X-ray weak with no clear evidence of
\xray\ obscuration.
X-ray weakness caused by X-ray obscuration is usually identified from 
an X-ray spectral shape that is 
flatter than
the intrinsic $\Gamma\approx2$ power law,
as soft \hbox{X-ray} photons are more heavily absorbed
than hard X-ray photons.
The effective power-law photon index ($\Gamma_{\rm eff}$) of an obscured
spectrum should be smaller than 2; in case of heavy absorption, the 0.5--5 keV 
$\Gamma_{\rm eff}$ can even reach a negative value. However, due to the frequent appearance
of partial-covering absorption 
in AGN X-ray spectra \citep[e.g.,][]{Immler2003,Ricci2017,Leighly2019} 
with a small fraction of the 
intrinsic coronal emission leaking through the absorber, a soft X-ray \hbox{($\lesssim5$~keV)} 
spectrum alone might
be insufficient for identifying heavy ($N_{\rm H}\gtrsim5\times10^{23}$~cm$^{-2}$) 
or Compton-thick ($N_{\rm H}>1.5\times10^{24}$~cm$^{-2}$) 
\xray\ obscuration, as the X-ray emission could be extremely weak
(e.g., $\Delta\alpha_{\rm OX}<-0.3838$ or X-ray weakness factor of $>10$)
yet the spectral shape is nominal with $\Gamma_{\rm eff}\approx2$, dominated by the leaked component (e.g., see Section 4.2 below for illustration).
On the other hand, the hard \xray\ \hbox{($\gtrsim5$~keV)} spectrum in this case should
generally
still be flat with $\Gamma_{\rm eff}\approx0\textrm{--}1$, as the Compton-reflection component,
from either the absorber or other reflectors (e.g., disk or torus), is 
expected to dominate
\citep[e.g.,][]{George1991,Comastri2011,Gandhi2014,Rovilos2014}.
It is based on these arguments that a small number of quasars have been proposed to be intrinsically
\xray\ weak; they are significantly X-ray weak yet with nominal ($\Gamma_{\rm eff}\approx2$) 
hard X-ray spectral shapes. 
The low-redshift ($z\lesssim1$) candidates include
a few BAL quasars that have \nustar\ observations \citep{Luo2013,Luo2014}, and
the high-redshift (\hbox{$z\approx1.5$--3}) candidates include a few
Large Bright Quasar Survey (LBQS) BAL quasars with \chandra\ observations \citep{Liu2018} and
a few luminous quasars with \xmm\ observations \citep{Nardini2019}.
Given the significant X-ray weakness, the \xray\ spectra of these candidates have very limited photon
statistics, and thus the spectral shapes are poorly constrained and sometimes even rely upon 
stacking analysis to obtain information on their average spectral properties
\citep[e.g.,][]{Luo2014}.

One quasar that is considered a prototypical example of intrinsically X-ray weak quasars
is PHL~1811,
a non-BAL quasar at $z = 0.192$ \citep{Leighly2007}.
It is
\hbox{X-ray} weak by a factor of \hbox{$\approx40$--130} with a steep 0.3--5~keV \xmm\
spectrum ($\Gamma\approx2.3$). There are no published hard X-ray
constraints, but its fast \hbox{X-ray} variability (flux varying
by a factor of $\approx4$ in 12 days) argues against
a heavily obscured spectrum dominated by distant reflection/scattering \citep{Leighly2007-2,Leighly2007}.
PHL~1811 has distinctive UV emission-line properties, including a small \iona{C}{iv}
rest-frame equivalent width (REW), a large \iona{C}{iv} 
blueshift, and strong UV
\iona{Fe}{ii} and \iona{Fe}{iii} emission.
\citet{Luo2015} conducted a systematic survey of the X-ray properties of 
PHL~1811 analogs, a sample of high-redshift quasars selected with 
UV emission-line properties similar to those of PHL~1811.
This sample of 18 PHL~1811 analogs turned out to be almost exclusively (17/18; 94\%)
X-ray weak, by an average X-ray weakness factor of 39. However, unlike
PHL~1811 itself, many 
of these PHL~1811 analogs appear to be X-ray obscured, as the sample stacked 
effective photon index 
($\Gamma_{\rm eff}=1.16_{-0.32}^{+0.37}$) indicates a flat spectral shape on average.
\citet{Luo2015} speculated that Occam's razor would
favor a uniform explanation for the \hbox{X-ray} weakness of all these
objects and that perhaps hard X-ray data of PHL~1811 would reveal a highly obscured component.

In this paper, we present improved hard X-ray constraints from 
deeper \nustar\ observations of 
two low-redshift intrinsically \hbox{X-ray} weak quasar
candidates in \citet{Luo2014}, 
PG~$1001+054$ (hereafter PG~1001) 
and PG~$1254+047$ (hereafter PG~1254).
We also provide for the first time hard \hbox{X-ray} constraints for 
PHL~1811 using archival \nustar\ and \xmm\ observations.
Based on the results, we suggest that X-ray obscuration is present in all
three quasars, and we propose that intrinsic X-ray weakness is not required 
to explain the X-ray weakness of these objects. The \nustar, \chandra, and \xmm\ data
of these quasars can be uniformly explained with an obscuration scenario
where the partial-covering absorber has variable column density and leaked fraction (partial-covering fraction).
The paper is organized as follows.
We describe the soft and hard \hbox{X-ray} observations and data analyses in Section~2.
We present \xray\ and multiwavelength properties of the three quasars in Section~3.
In Section~4, we propose that these quasars are likely X-ray obscured and describe how
an obscuration scenario may explain their soft and hard \hbox{X-ray} weakness, without
invoking intrinsic \hbox{X-ray} weakness. 
The absorber is likely the clumpy dust-free wind launched from the accretion disk.
We summarize in Section~5. In the Appendix, we present new hard \xray\ constraints
from a recent \xmm\ observation of LBQS~$1442-0011$, a high-redshift 
intrinsically \hbox{X-ray} weak quasar
candidate in \citet{Liu2018}; the results are consistent with our proposed obscuration
scenario.

Throughout this paper, we use a cosmology with $H_0=67.4$~km~s$^{-1}$~Mpc$^{-1}$,
$\Omega_{\rm M}=0.315$, and $\Omega_{\Lambda}=0.685$ \citep{Planck2020}. Uncertainties are
quoted at a 1$\sigma$ confidence level, and limits are at a $90\%$ confidence level.
The energy ranges used in our photometric and spectroscopic analyses are 0.3--8~keV
for \chandra\ observations, 0.3--10~keV for \xmm\ observations, and 3--24~keV
for \nustar\ observations, unless otherwise specified.
Due to the X-ray weakness of the three quasars, all X-ray spectra 
were grouped with at least one count per bin, and the 
W statistic\footnote{\url{
https://heasarc.gsfc.nasa.gov/docs/xanadu/xspec/manual/XSappendixStatistics.html.} \label{wstat}} of XSPEC 
(v12.10.1; \citealt{Arnaud1996}) was used in spectral fitting. Galactic absorption
\citep{HI4PI2016} was included
in the \xray\ spectral modeling.

\section{Object Properties, X-ray Observations, and X-ray Data Analyses}

\begin{deluxetable*}{lcccccccrrr}
\tablecaption{Basic Object Properties and List of Observations}
\tablehead{
\colhead{Object }  &
\colhead{$z$}  &
\colhead{$m_{B}$}  &
\colhead{$\log M_{\rm BH}$}  &
\colhead{$L/L_{\rm Edd}$} &
\colhead{$\log L_{\rm 2500~{\textup{\AA}}}$} &
\colhead{$N_{\rm H, Gal}$}&
\colhead{Observatory}	&
\colhead{Observation}	&
\colhead{Obs. Date} &
\colhead{Exp} \\
\colhead{Name}	&
\colhead{}	&
\colhead{}	&
\colhead{($M_{\Sun}$)}	&
\colhead{}	&
\colhead{(erg s$^{-1}$ Hz$^{-1}$)}	&
\colhead{($10^{20}$~cm$^{-2}$)}	&
\colhead{} &
\colhead{ID}	&
\colhead{} &
\colhead{(ks)}\\
\colhead{(1)}         &
\colhead{(2)}         &
\colhead{(3)}         &
\colhead{(4)}         &
\colhead{(5)}         &
\colhead{(6)}         &
\colhead{(7)}         &
\colhead{(8)}		  &
\colhead{(9)}		  &
\colhead{(10)}		  &
\colhead{(11)}		 
}
\startdata
PG~1001+054	&0.161	&16.1   &7.74	&0.5	  &  29.9& 2.38&\xmm	&0150610101	&2003 May 4  &8.6	\\
		&		&		&		&	&	&	&\chandra	&11852	&  2010 Jan 11	  &4.9		\\
		&		&		&		&	&	&	&\nustar		&60001122002	& 2013 Jun 28	      &19.6	\\
		&		&		&		&	&	&	&\nustar		&60501014001& 2020 May 23		&101.1\\
PG~1254+047 &1.026	&15.8  &9.68	&0.3	&31.5&2.03  	&\chandra	&832	 &2000 May 29	&36.0	\\
		&		&		&		&	&	&	&\nustar 	&60001123002	& 2013 Jun 27	     &29.4	\\
		&		&		&		&	&	&	&\nustar 	&60401013002	& 2019 Jun 8	    &88.0	\\
PHL~1811 &0.192	&13.9   &8.25	&1.6	&30.9&4.22  	&\chandra 	&2975		&   2001 Dec 5 &9.3		\\
		 &		&		&		&	&	&	&\chandra 	& 2985	& 2001 Dec 17     &9.8		\\
		 &		&		&		&	&	&	&\xmm 		&0204310101		& 2004 Nov 1    &23.5	\\
		 &		&		&		&	&	&	&\xmm 		&0761910201	& 2015 Nov 29	       &39.9	\\
		 &		&		&		&	&	&	&\nustar 	&60101004002	& 2015 Nov 28	      &54.7
\enddata

\tablecomments{
Cols. (1) and (2): object name and redshift.
Col. (3): $B$--band magnitude.
Cols. (4) and (5): single-epoch virial BH mass and Eddington ratio 
from \citealt{shen2011} (for PG~1001 and PG~1254) or \citealt{Leighly2007} 
(for PHL~1811).
Col. (6): 2500~{\textup{\AA}} monochromatic luminosity from \citealt{shen2011} (for PG~1001 and PG~1254) or \citealt{Leighly2007} (for PHL~1811).
Col. (7): Galactic neutral hydrogen column density \citep{HI4PI2016}.
Cols. (8) and (9): observatory and observation ID.
Col. (10): observation start date.
Col. (11): cleaned exposure time; for \nustar\ observations, 
it is the average value of the FPMA and FPMB exposure times.
}
\end{deluxetable*}

\subsection{Basic Object Properties}
The basic properties of the three quasars are listed in
Table~1.
PG~1001
is at $z = 0.161$ with a $B$-band magnitude of 16.1. 
Its H$\beta$-based single-epoch virial SMBH mass ($M_{\rm BH}$)
is $\approx5.5\times10^{7} M_{\Sun}$, and the estimated
Eddington ratio is $\approx0.5$ \citep{shen2011}.
The full width at half maximum (FWHM) of its H$\beta$ emission line is 1740~km~s$^{-1}$
and thus it was classified as a narrow-line 
type 1 quasar (NLQ1; \citealt{Wills2000}), which refers to type 1 
quasars with  
narrow (FWHM$<2000$~km~s$^{-1}$) 
H$\beta$ emission lines (e.g., Footnote 1
of \citealt{Wills2000}).
It also shows weaker [\iona{O}{iii}]~$\lambda5007$ emission compared to 
typical quasars (REW 
smaller by a factor of $\approx2$; \citealt{Vandenberk2001,shen2011}).

 PG~1254 is a luminous quasar at $z = 1.026$ with a 
 $B$-band magnitude of 15.8. The \iona{Mg}{ii}-based single-epoch virial SMBH mass 
is $\approx4.8\times10^{9} M_{\Sun}$, 
 and the estimated
 Eddington ratio is $\approx0.3$ \citep{shen2011}. 
There are no H$\beta$ or [\iona{O}{iii}] measurements available in the literature. 


PHL~1811, at $z = 0.192$, is very bright with a $B$-band magnitude 
of 13.9. Its H$\beta$-based 
single-epoch virial SMBH mass is $\approx1.8\times10^{8} M_{\Sun}$, and 
its estimated Eddington ratio is $\approx1.6$ \citep{Leighly2007}. 
We caution that the virial SMBH mass and the estimated Eddington ratio
are subject to large uncertainties,
especially in the super-Eddington regime, as the virialization
assumption may no longer be valid when the
broad emission-line region (BELR) is likely exposed to
large and anisotropic radiation pressure \citep[e.g.,][]{Marconi2008,Marconi2009,
Netzer2010}.
PHL~1811 was classified as a NLQ1 with no apparent [\iona{O}{iii}]~$\lambda5007$ emission
\citep{Leighly2007-2}.

\subsection{\nustar\ Observations and Data Analysis}
PG~1001 and PG~1254 were observed by \nustar\ in 2013 with 
cleaned exposure times of 19.6~ks and 29.4~ks, 
respectively. They were not detected in the 8--24 keV band individually, but 
they were considered good candidates for intrinsically X-ray weak quasars 
from X-ray stacking analysis \citep{Luo2014}.
We obtained much deeper \nustar\ observations of 
PG~1254 in 2019 and PG~1001 in 2020 with exposure times 
of $\approx 100$~ks {(PI: W.~N.~Brandt)}, with the aim of detecting them individually in the 8--24 keV band and 
providing improved spectral-shape constraints.
PHL~1811 has an archival \nustar\ observation
in 2015 with a cleaned exposure time of 54.7~ks, simultaneous to
a 39.9~ks \xmm\ observation (PI: K.~Leighly). 
The details of the 
\nustar\ observations are listed in Table~1. 

For \nustar\
data reduction, we used HEASoft (v6.29) and the \nustar\ Data 
Analysis Software (NuSTARDAS; v2.1.1) with \nustar\ CALDB 20210728. 
We used the {\sc nupipeline} script to generate calibrated clean 
event files from the unfiltered event files of the two 
focal plane module detectors (FPMA and FPMB). 
For the 2019 \nustar\ observation of PG~1254, 
background event rates were slightly elevated around the South Atlantic Anomaly (SAA),
and we followed the recommendations from the 
\nustar\ instrument team and applied an additional SAA filter
({\sc saamode} = {\sc optimized},
{\sc tentable} = {\sc yes}, and {\sc saacalc} = 1)
during the {\sc nupipeline} processing.
We created \nustar\ images using the Chandra Interactive Analysis 
of Observation (CIAO; v4.13)\footnote{http://cxc.harvard.edu/ciao/.}
tool {\sc dmcopy} in three energy bands: \hbox{3--24~keV} (full band), 
3--8~keV (soft band), and 8--24~keV (hard band).

For each \nustar\ observation, we co-added the FPMA and FPMB 
images to 
improve sensitivity, which helps in detecting
and characterizing faint \nustar\ sources \citep[e.g.,][]{Lansbury2017}. 
The co-added images were 
created by combining the FPMA and FPMB images in each of 
the three bands using the HEAsoft tool {\sc ximage}. These images 
were then used for source detection and 
aperture photometry.
For each co-added image, we searched for \hbox{X-ray} 
sources using the CIAO tool {\sc wavdetect} \citep{Freeman2002} 
with a false-positive probability threshold of $10^{-5}$ and 
wavelet scales of 2, 2.83, 4, 5.66, 8, 11.31, and 16 pixels \citep[e.g.,][]{Luo2014};
the \nustar\ pixel size is $2.46\arcsec$. 
In the 2013 observations, PG~1001 and PG~1254
were detected in the soft and full bands but not in the hard band.
They were detected in all three bands in the latest deeper observations. 
PHL~1811 was also detected in all three bands. 
In the following analysis, we 
adopted
the full-band {\sc wavdetect} positions as the X-ray positions. 
The offsets between the X-ray and optical positions range from 
$3.4\arcsec$ to $9.4\arcsec$, with are
typical for faint \nustar\ sources \citep[e.g.,][]{Lansbury2017}.

We extracted source and background spectra using the {\sc nupipeline} script.
We used $35\arcsec$-radius circular source regions centered on the
X-ray positions, and annular background regions centered on the
X-ray positions with inner radii of $120\arcsec$
and outer radii of $180\arcsec$.
We verified that there are no sources
in the background regions.
For each observation, we merged the FPMA and FPMB source spectra,
background spectra, and response files using the HEASoft tool
{\sc addspec}.

\subsection{Archival \chandra\ and \xmm\ Observations and Data Analyses}

The archival \chandra\ and \xmm\ observations of the three 
quasars are listed in 
Table~1.\footnote{PHL 1811 has another \chandra\ observation in 2012 with an
exposure time of 2.0~ks (see Footnote 18 of \citealt{Luo2015});
we do not use this short observation in
the present study as it does not have sufficient statistics 
to place meaningful constraints on relevant parameters.}
For \chandra\ data reduction,
we used the CIAO {\sc chandra\_repro} script to generate new level 2 
event files. Background flares were then filtered with the 
{\sc deflare} script using an iterative 3$\sigma$ clipping 
algorithm. We created 
0.5--8~keV images from the cleaned event files using {\sc dmcopy}. 
We searched for \hbox{X-ray} sources in the \hbox{X-ray} images using 
{\sc wavdetect} with a false-positive 
probability threshold of $10^{-6}$ and wavelet scales of 1, 1.414, 2, 
2.828, 4, 5.656, and 8 pixels. The three quasars were all 
detected within $0.22\arcsec$--$0.42\arcsec$ of the optical positions. 
For each observation, the source spectrum was extracted 
using the {\sc specextract} tool
from a circular region centered on the \hbox{X-ray} position with a radius 
of $2\arcsec$. 
The background spectrum was extracted from an annular region centered 
on the \hbox{X-ray} position with an inner radius of 6\arcsec\ and an outer radius of 10\arcsec; we verified that the background regions do not contain any X-ray sources.

For the \xmm\ observations, we used only the data from 
the pn camera.\footnote{We have checked the MOS data, which have lower 
photon statistics, especially in the 5--10~keV band that is of interest to this study;
combining 
the pn spectra with the low signal-to-noise ratio MOS spectra 
might introduce additional systematic uncertainties.}
We used the Science Analysis System (SAS; v1.2) 
to process the data, following the standard procedure described 
in the SAS Data Analysis Threads.\footnote{http://www.cosmos.esa.int/web/xmm-newton/sas-threads.}
We used the {\sc epproc} tool to produce calibrated event files. 
A threshold of 0.4~cts~s$^{-1}$ was adopted to filter 
background flares. We created good-time-interval files 
using the {\sc tabgtigen} script, and we generated cleaned event 
files using the {\sc evselect} tool. For each observation, we used 
the {\sc evselect} tool to extract source and background spectra, with a  
$30\arcsec$-radius circular source region centered on the
optical position of the quasar and a $80\arcsec$-radius
circular, source-free background region on the same CCD chip as the source region.

For the 2015 \xmm\ observation of PHL~1811, 
we also used the Optical Monitor (OM) photometric data for constructing its 
SED.\footnote{The 2004 OM observation of PHL~1811 
used only one filter and its photometric measurement 
does not suggest
any variability compared to the 2015 OM results.} 
We generated nine exposures for the five filters 
(UVW2, UVM2, UVW1, U, B) using the {\sc omichain} script, 
and the photometric measurements of every exposure were recorded in the 
SWSRLI files. We extracted the magnitude measurements from these files 
and adopted the mean magnitude for each filter. 
The coincidence loss corrections (due to the high count rates) 
were applied during the pipeline 
processing.

\section{\hbox{X-ray} and Multiwavelength Properties}

\subsection{Soft \hbox{X-ray} Weakness and Obscuration Signatures from Archival 
X-ray Observations}
\subsubsection{PG~1001}
PG~1001 was observed by \xmm\ and \chandra\ in 2003 and 2010 with exposure
times of 8.6~ks and 4.9~ks, respectively, and the results were presented in
\citet{Schartel2005} and \citet{Saez2012}.
As described in Section 2.3, 
we reduced these data and extracted the corresponding \xray\ spectra.
We performed simple power-law spectral fitting of the
2003 \xmm\ spectrum in the 1--10~keV band
and the 2010 \chandra\ spectrum in the \hbox{1--8~keV} band; a lower energy bound of 1~keV
was adopted here because there is apparent soft X-ray excess emission
in the 0.3--1~keV \xmm\
spectrum that is probably related to ionized
absorption \citep{Schartel2005}.
The resulting power-law photon
indices are $0.8\pm0.2$ and $1.3^{+1.1}_{-1.0}$, indicative of \xray\ obscuration.
From the best-fit results, we derived two $\alpha_{\rm OX}$ values for PG~1001, and
they are shown in the $\alpha_{\rm OX}$
versus $L_{\rm 2500~{\textup{\AA}}}$ plane in Figure~1;
the $L_{\rm 2500~{\textup{\AA}}}$ measurement
is from \citet{shen2011}. Besides the soft X-ray weakness, there is also 
X-ray flux variability between these two observations. 
We then computed the $\Delta\alpha_{\rm OX}$ parameters,
which are $-0.58\pm0.15$ for the \xmm\ observation and $-0.75\pm0.16$
for the \chandra\ observation, corresponding to X-ray weakness factors
of $32^{+47}_{-19}$ and $88^{+144}_{-55}$ at rest-frame 2~keV, respectively.
The $\Delta\alpha_{\rm OX}$ uncertainties were dominated by
the $\alpha_{\rm OX}$ rms scatter ($\approx0.15$; Table~5 of
\citealt{steffen2006})
of the $\alpha_{\rm OX}$--$L_{\rm 2500~{\textup{\AA}}}$ relation.
We note that
there does not appear to be any
strong long-term UV/optical variability of the three quasars in our study
(maximum flux variability amplitudes 
$\approx16\%$--80\%; see Section~3.3 below), and thus the large X-ray weakness factors
derived using non-simultaneous
\xray\ and UV/optical data are not significantly affected by any UV/optical variability.

\subsubsection{PG~1254}
PG~1254 has a 36.0~ks \chandra\ observation in 2000 that was studied in \citet{Sabra2001}.
We performed simple power-law spectral fitting of this \chandra\ spectrum, and the
resulting photon index is $0.6\pm0.3$, indicative
of X-ray obscuration. Its location in the $\alpha_{\rm OX}$
versus $L_{\rm 2500~{\textup{\AA}}}$ plane is shown in Figure~1;
the $L_{\rm 2500~{\textup{\AA}}}$ measurement
is from \citet{shen2011}. The corresponding $\Delta\alpha_{\rm OX}$ value is $-0.74\pm0.16$,
indicating an X-ray weakness factor of $87^{+139}_{-53}$ at \hbox{rest-frame} 2~keV.

\subsubsection{PHL~1811}
PHL 1811 has two \chandra\ observations and two \xmm\ observations, as listed in
Table~1; the first three were presented in \citet{Leighly2007}.
From simple power-law spectral fitting of the two \chandra\ and two \xmm\
spectra, we obtained steep spectral shapes with photon indices in the range of $\approx$2.0--2.6,
consistent with those in \citet{Leighly2007}.
The two $\alpha_{\rm OX}$ values from the 2001 December 5 and December 17 \chandra\
observations are shown in Figure 1, using the
$L_{\rm 2500~{\textup{\AA}}}$ value adopted from \citet{Leighly2007}.
The $\Delta\alpha_{\rm OX}$ values are $-0.78\pm0.15$ for the December 5 observation
and $-0.58\pm0.15$ for the December 17 observation, corresponding to X-ray weakness
factors of $108^{+160}_{-64}$ and $33^{+48}_{-20}$ at rest-frame 2~keV, respectively.
We do not show the $\alpha_{\rm OX}$ values from the two
\xmm\ observations in Figure~1, as they are
very close to the 2001 December 5 \chandra\ value (e.g., $\Delta\alpha_{\rm OX}=-0.81$ for
the 2015 \xmm\ observation).

In summary, previous soft X-ray observations of the three quasars have revealed
significant soft X-ray weakness, with the rest-frame 2~keV fluxes
$\approx32$--108 times weaker compared to the expectations from their optical/UV emission.
The soft X-ray spectra of PG~1001 and PG~1254 have
flat spectral shapes, indicative of X-ray obscuration, while
the soft X-ray spectra of PHL~1811 do not show evidence for X-ray obscuration.

\begin{figure}
\centering
\includegraphics[trim=0 0 20 20,clip, width=1\linewidth]{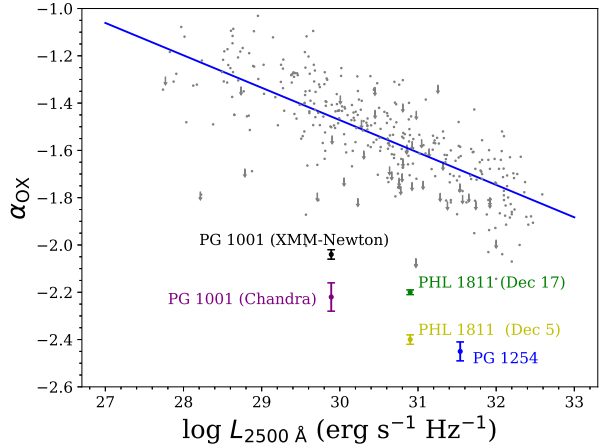}
\caption{
X-ray-to-optical power-law slope parameter ($\alpha_{\rm OX}$)
vs.\ 2500~\AA\ monochromatic luminosity for the three quasars,
showing their significant soft X-ray weakness derived using archival 
\chandra\ and \xmm\ observations.
For PHL~1811, the $\alpha_{\rm OX}$ values from the 2001 December 5 and December 17 \chandra\
observations are shown; the $\alpha_{\rm OX}$ measurements from the two
\xmm\ observations not shown here are
very close to the 2001 December 5 \chandra\ value.
The solid blue
line represents the $\alpha_{\rm OX}$--$L_{\rm 2500~{\textup{\AA}}}$
relation from \citet{steffen2006}. The small gray dots and downward
arrows represent the $\alpha_{\rm OX}$ values and upper limits of the
\citet{steffen2006} AGN sample, respectively.
}
\end{figure}

\subsection{Hard \hbox{X-ray} Weakness and Spectral Shape Constraints from \nustar}

For each \nustar\ observation, we performed aperture photometry using the co-added 
FPMA $+$ FPMB images in the three bands (soft, hard, and full). In each image, 
we extracted source counts ($S$) and background
counts ($B$) from the same source and background regions
used in the spectral extraction in Section 2.2. The encircled-energy fraction of the 
source region is $63.9\%$ according to the \nustar\ point-spread function. 
We determined the source significance 
by calculating the binomial no-source probability ($P_{\rm B}$; 
e.g., \citealt{Luo2013}), which is defined as
\begin{equation}
P_{\rm B}(X\ge S)=\sum_{X=S}^{N}\frac{N!}{X!(N-X)!}p^X(1-p)^{N-X}~.
\end{equation}
In this expression, $N$ = $S+B$ and $p = 1/(1 +BACKSCAL$), 
where $BACKSCAL$ is the ratio between the exposure-time weighted 
areas of the background and source regions. A smaller $P_{\rm B}$ value 
indicates a more significant signal. We considered the source detected 
in a band if the measured $P_{\rm B}$ value is smaller than 0.01 
(corresponding to a $>2.6\sigma$ significance level). 
With this criterion, PG~1001 and PG~1254 were not detected in 
the hard band in their 2013 observations, and the three quasars 
were detected in all the other images; these results are consistent with the 
{\sc wavdetect} results in Section 2.2. 
In the hard band, the $P_{\rm B}$ values for PG 1001 and PG 1254 in their latest
observations are $5.4\times10^{-7}$ ($5.0\sigma$) and $3.8\times10^{-6}$ ($4.6\sigma$),
respectively, and $P_{\rm B}$ is $1.1\times10^{-3}$ ($3.3\sigma$) for PHL~1811; 
these values indicate significant detections in the hard band.
The \nustar\ hard-band images of the there quasars are displayed in Figure~2.

For the detected sources, 
we computed their \hbox{aperture-corrected} net counts $(S-B/BACKSCAL)/0.639$. 
The associated errors were derived from the $1\sigma$ Poisson errors of the extracted source 
and background counts \citep{Gehrels1986}. 
Compared to PG~1001 and PG~1254, PHL~1811 has larger relative 
count errors in the hard band, consistent with its larger hard-band $P_{\rm B}$ value 
(lower detection significance).
For undetected sources, 
we calculated $90\%$ confidence-level upper limits on the source counts 
following the Bayesian approach of \citet{Kraft1991}. The net counts 
and upper limits in the three bands are listed in Table~2.

\begin{figure}
\centering
\includegraphics[trim=100 150 60 180,clip, width=1.0\linewidth]{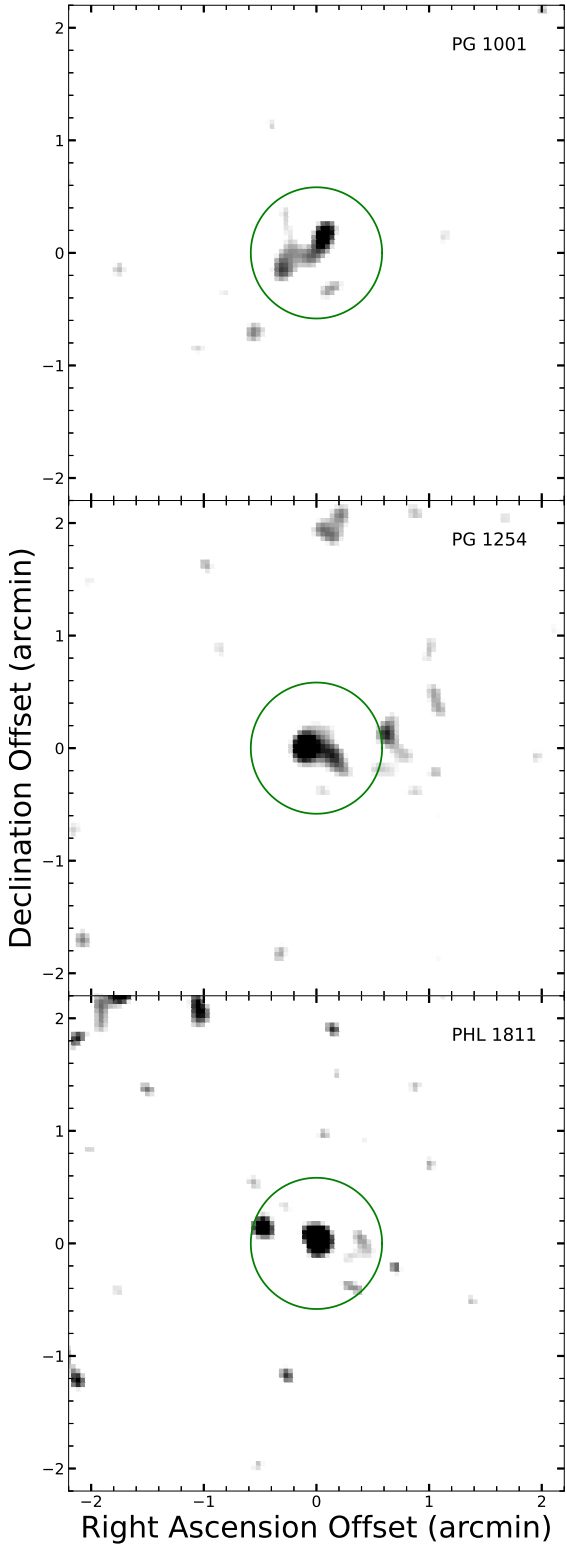}
\caption{\nustar\ hard-band (8--24 keV) images of the three quasars, 
smoothed with a Gaussian kernel with standard deviation of 2.0 pixels. 
Each image is centered on the corresponding X-ray source position, and the circle 
indicates
the $35\arcsec$-radius aperture used for source extraction.
Due to statistical fluctuations, the smoothed images of these
faint sources do not have circular 
morphologies following the \nustar\ on-axis point spread function.
}
\end{figure}

 For each quasar in each observation, we derived an effective power-law 
 photon index ($\Gamma_{\rm eff}$) from the band ratio, which 
 is the ratio between the hard-band (8--24~keV)
 and soft-band (3--8~keV) counts, based on the following procedure.
 (1) For a given set of $\Gamma$ values, we produced a set of mock power-law spectra
 using the XSPEC {\sc fakeit} 
 routine and the spectral response files. (2) For each mock spectrum, we computed the corresponding 
 band ratio. (3) We interpolated the $\Gamma$ versus band ratio set 
 to derive the $\Gamma_{\rm eff}$ value from the measured band ratio.
The $\Gamma_{\rm eff}$ values are listed in Table~2.
The $1\sigma$ errors on $\Gamma_{\rm eff}$ were propagated from 
 the errors of the band ratios derived using {\sc behr} \citep{Park2006}. 
 If the	quasar was not detected in the hard band, we computed the lower 
 limit on $\Gamma_{\rm eff}$ from the upper limit on the band ratio 
 calculated using {\sc behr}. In this case, $\Gamma_{\rm eff} = 2.0$ was 
 adopted in the following calculations of fluxes, flux densities, 
 and luminosities; we consider this more appropriate than using the lower limit value,
and adopting a value different from $2.0$ would not affect the results significantly.

To compute fluxes, 
we obtained 
conversion factors from count rates to fluxes in the three 
bands using the mock spectrum with a photon index of $\Gamma_{\rm eff}$. 
Flux errors were propagated from the count errors, and flux upper 
limits were derived from the upper limits on the net counts. 
The luminosity in the rest-frame 2--10~keV band ($L_{\rm X}$) in each observation 
was derived from the 
full-band flux adopting a power-law spectrum with a photon index 
of $\Gamma_{\rm eff}$. 
The X-ray fluxes and luminosities
are listed in Table~2.
For PG~1001 and PG~1254 that have two \nustar\ observations,
the photometric properties from the two observations are overall consistent
within the errors, except for
the $\Gamma_{\rm eff}$ constraints of PG~1001 which suggest possible
spectral-shape evolution.
We stacked their photometric measurements
to derive average properties; the results are also listed in Table~2.

We calculated the factor of \hbox{X-ray} 
weakness at rest-frame 8~keV ($f_{\rm w}$), which is defined as the ratio between 
the expected and observed 8~keV flux density 
{($f_{\rm w}={f_{\rm \nu,~expected}}/{f_{\rm \nu,~observed}}$).}
The observed 8~keV flux density was computed from the full-band flux for a power-law 
spectrum with a photon index of $\Gamma_{\rm eff}$, and the expected 
8~keV flux density was calculated from the \citet{steffen2006} 
$\alpha_{\rm OX}$--$L_{\rm 2500~{\textup{\AA}}}$ relation 
adopting a $\Gamma=2$ power-law spectrum. 
The $f_{\rm w}$ uncertainties were propagated from
the $\alpha_{\rm OX}$ rms scatter 
of the $\alpha_{\rm OX}$--$L_{\rm 2500~{\textup{\AA}}}$ relation.
{We caution that $f_{\rm w}$ is different from the factor of X-ray weakness
quantified by the $\Delta\alpha_{\rm OX}$ parameter (Section 1), which 
is for rest-frame 2~keV.}

Compared to the previous hard-band non-detections, 
the deeper \nustar\ observations of PG~1001 and PG~1254 improve the source-detection
significance and provide better constraints on the hard X-ray weakness factors and 
hard \xray\ spectral shapes.
The \nustar\ observation of PHL~1811 provides first-ever hard X-ray constraints for this 
extreme quasar.
PG~1001 and PHL~1811 are significantly hard X-ray weak
($f_{\rm w}\approx26$--74), while PG~1254 is only \xray\ weak
by a factor of $\approx2.7$ in the hard X-rays 
despite its significant soft X-ray weakness (Figure~1).
Although the 2013 observation of PG~1001 suggests
a potentially typical hard X-ray spectral shape 
($\Gamma_{\rm eff}>1.5$), the 2020 observation and the $2013+2020$ stacked data 
suggest flat 
spectral shapes ($\Gamma_{\rm eff}\approx0.4$--1.0). 
Both observations of PG~1254 suggest
nominal spectral shapes with 
$\Gamma_{\rm eff}\approx1.8$. The PHL~1811 observation surprisingly reveals
that 
its hard X-ray photon index ($\Gamma_{\rm eff}=1.4_{-0.7}^{+0.8}$),
though loosely constrained, appears marginally smaller than 
its soft X-ray (0.3--5 keV) photon index ($2.3\pm0.1$) from the 2004 \xmm\ observation 
\citep{Leighly2007}.

\begin{deluxetable*}{lccccccccccc}
\tablecaption{\nustar\ Photometric Properties}
\tablehead{
\colhead{Object Name }  &
\colhead{Obs. Year}  &
\multicolumn{3}{c}{Net Source Counts$^{a}$} &
\colhead{$\Gamma_{\rm eff}$$^{b}$} &
\multicolumn{3}{c}{Flux ($10^{-14}$~\flux)} &
\colhead{$\log L_{\rm X}$ (erg s$^{-1})$}	&
\colhead{$f_{\rm w}$$^{c}$} \\
\cline{3-5}  \cline{7-9}
\colhead{}	&
\colhead{}	&
\colhead{3--24}                   &
\colhead{3--8}                   &
\colhead{8--24}                   &
\colhead{}                  &
\colhead{3--24}                   &
\colhead{3--8}                   &
\colhead{8--24}                   &  
\colhead{2--10}	&
\colhead{} \\
\colhead{}  &
\colhead{}  &
\colhead{keV}  &
\colhead{keV}  &
\colhead{keV}  &
\colhead{}  &
\colhead{keV}  &
\colhead{keV}  &
\colhead{keV}  &
\colhead{keV}  &
\colhead{}  &
}
\startdata
PG~1001 &2013 &$                    52^{+16}_{-15}$&$40^{+13}_{-11}$&$                    <26$&$>1.5~(2.0)$&$        7.8^{+2.4}_{-2.2}$&$        4.4^{+1.2}_{-1.4}$&$        <5.9$& $42.7$&   $16^{+24}_{-10}$\\
PG~1001 &2020 &$                    172^{+35}_{-34}$&$57^{+24}_{-23}$&$                    113^{+27}_{-25}$&$ 0.4^{+0.6}_{-0.9}$&$        7.8^{+1.6}_{-1.5}$&$        1.1^{+0.5}_{-0.4}$&$6.6^{+1.6}_{-1.5}$&   $42.0$&   $34^{+50}_{-20}$\\
PG~1001 &2013+2020&$                223^{+38}_{-37}$&$98^{+26}_{-25}$&$125^{+29}_{-27}$&$ 1.0^{+0.5}_{-0.6}$&$7.2\pm1.2$&$1.7^{+0.5}_{-0.4}$&$5.5^{+1.3}_{-1.2}$&$42.2$&$26^{+38}_{-15}$\\
\\
PG~1254 &2013 &$                    63^{+21}_{-19}$&$35^{+15}_{-14}$&$                    <50$&$>0.4~ (2.0)$&$        6.2^{+2.0}_{-1.9}$&$        2.5^{+1.1}_{-1.0}$&$        <7.3$&    $44.5$&   $3.4^{+5.0}_{-2.0}$\\
PG~1254 &2019 &$                    275^{+38}_{-36}$&$171^{+28}_{-26}$&$                    104^{+27}_{-25}$&$ 1.8^{+0.5}_{-0.4}$&$        10.0^{+1.4}_{-1.3}$&$        4.3\pm0.7$&$ 5.7^{+1.5}_{-1.4}$&    $44.5$&   $2.5^{+3.6}_{-1.5}$\\
PG~1254 &2013+2019&$                338^{+43}_{-41}$&$       206^{+31}_{-30}$&$                132^{+30}_{-29}$&$ 1.8\pm0.3$&$ 9.1^{+1.2}_{-1.1}$&$ 3.8\pm0.6$&$ 5.3\pm1.2$&    $44.5$&$2.7^{+3.9}_{-1.6}$\\
\\
PHL~1811 &  2015&$                    113^{+28}_{-26}$&$                    59^{+20}_{-18}$&$                    55^{+21}_{-19}$&$1.4^{+0.8}_{-0.7}$&$        7.0^{+1.7}_{-1.6}$&$        2.2^{+0.8}_{-0.7}$&$4.9^{+1.8}_{-1.7}$&   $42.6$    &   $74^{+109}_{-44}$
\enddata

\tablenotetext{a}{{The errors were derived from the 1$\sigma$ errors of the 
extracted source and background counts \citep{Gehrels1986}.
For undetected sources, we calculated $90\%$ confidence-level upper limits on the
source counts following the Bayesian approach of \citet{Kraft1991}.}}
\tablenotetext{b}{Effective power-law photon index. If the source 
is not detected in the 8--24~keV band, a lower limit value is provided, but 
$\Gamma_{\rm eff} = 2.0$ 
(as shown in parentheses) was adopted in calculating the fluxes, 
flux densities, and luminosity.}
\tablenotetext{c}{Factor of X-ray weakness at rest-frame 
8 keV, derived by comparing the observed 8 keV flux density to that 
expected from the \citet{steffen2006} 
$\alpha_{\rm OX}$--$L_{\rm 2500~{\textup{\AA}}}$ relation assuming 
a $\Gamma = 2$ power-law spectrum. The uncertainty is dominated by the scatter
of the $\alpha_{\rm OX}$--$L_{\rm 2500~{\textup{\AA}}}$ relation.}
\end{deluxetable*}

\subsection{Spectral Energy Distributions and Optical/Infrared Variability}

We constructed infrared (IR) to \hbox{X-ray} SEDs 
for the three quasars. We collected IR--UV photometric data from the 
{Wide-field Infrared Survey Explorer} ({WISE}; \citealt{Wright2010}), 
{ Near-Earth Object WISE} ({NEOWISE}; {\citealt{Mainzer2014}}), 
Two Micron All Sky Survey (2MASS; \citealt{Skrutskie2006}), Sloan Digital 
Sky Survey (SDSS; \citealt{York2000}), and {Galaxy Evolution Explorer} 
({GALEX}; \citealt{Martin2005}) catalogs. 
For the two PG quasars, we also included their SED data from
\citet{Neugebauer1987}. We added Spitzer photometric measurements 
for PG 1001 \citep{Veilleux2009}.
For PHL~1811, we included the
2001 HST STIS UV spectrum and the 2015 \xmm\ OM measurements. The optical 
and UV data have been corrected for Galactic extinction following the de-reddening 
approach in \citet{Cardelli1989} and \citet{O'Donnell1994}. 
The SEDs are shown in Figure~3. For comparison, we also plotted in each
panel the mean SED of high-luminosity radio-quiet quasars in \citet{Krawczyk2013}
normalized to the 2500~\AA\ luminosity. 
The X-ray component of the mean quasar SED is a $\Gamma=2$ power-law continuum that follows 
the $\alpha_{\rm OX}$--$L_{\rm 2500~{\textup{\AA}}}$ relation.
The \hbox{IR-to-UV} SEDs of the
three quasars are broadly consistent with those of typical quasars;
{the slight deviations in the IR for PG~1001 and PHL~1811 are within
the scatters ($\approx0.2$--0.25 dex) 
of the mean quasar SED at these frequencies. }

We added soft and hard X-ray measurements to the SEDs.
We used the 2~keV luminosities 
from the power-law spectral fitting of the \chandra\ or \xmm\ 
spectra (Section~3.1). From the \nustar\ full-band fluxes (Section~3.2; stacked results
were used for PG~1001 and PG~1254),
we 
derived 8~keV and 15~keV luminosities 
adopting power-law spectra with the measured $\Gamma_{\rm eff}$ values. 
Compared to the typical quasar SED which follows the $\alpha_{\rm OX}$--$L_{\rm 2500~{\textup{\AA}}}$ relation,
the soft and hard X-ray weakness of the three quasars is 
evident. 
For PHL~1811, we also show the soft X-ray spectral slopes from the two \xmm\
observations and the hard \xray\ spectral slope
constrained from the \nustar\ observation. The spectral slopes differ beyond
the 1$\sigma$ level, suggesting that
X-ray obscuration might also be present in PHL~1811.
{The SEDs of the three quasars also show clearly that they deviate from
the observed $L_{\rm X}$--$L_{\rm MIR}$ relations for typical
quasars \citep[e.g.,][]{Lutz2004,Mateos2015,Stern2015,Chen2017,Martocchia2017}, with $L_{\rm MIR}$ being the mid-IR luminosity measured
at rest-frame 6~$\mu$m. The offsets from the relations are approximately quantified by 
the $f_{w}$ values (Table 2), and PHL~1811 is $\approx70$ times X-ray underluminous
compared to its mid-IR luminosity.}
Integrating the IR-to-X-ray SEDs, we estimate the bolometric 
luminosities to be $3.4\times10^{45}$~erg~s$^{-1}$ for PG~1001, 
$2.1\times10^{47}$~erg~s$^{-1}$ for PG~1254, and $4.2\times10^{46}$~erg~s$^{-1}$ 
for PHL~1811. These values are consistent with those provided in \citet{shen2011} and 
\citet{Leighly2007}.

The \hbox{IR-to-UV} SED data are not simultaneous and may be affected by
variability. Mild variability is apparent in the SED of PG~1001, where the
more recent optical and near-IR measurements are $\approx2$--60\% lower
than the \citet{Neugebauer1987} data.
To investigate the optical variability of these three quasars, we further 
examined their long-term optical light curves constructed using 
the public catalogs of the Zwicky Transient
Facility Data Release 9 (ZTF DR9; \citealt{Bellm2019}) and the
Catalina Real-Time Transient Survey (CRTS; \citealt{Drake2009}).
The maximum flux variability amplitudes in the ZTF $g$ band 
range from 13\% to 40\%, and in the
CRTS $V$ band they range from
16\% to 80\%; PG 1001 varied the most among the three quasars.
Since PG~1001 has the strongest optical variability, we 
also checked its IR light curve from the NEOWISE catalog.
Between 2014 May and 2020 November, its maximum flux variability amplitude
in the W1 (W2) band is 18\% (13\%).
The mild optical/IR variability observed in these quasars suggests that the overall
accretion power did not change significantly (e.g., by factors of $>2$) over the years.
Compared to the soft and hard X-ray weakness factors ($\Delta\alpha_{\rm OX}$
and $f_{\rm w}$ in Sections 3.1 and 3.2),
the UV/optical variability factors are
much smaller, indicating that the X-ray weakness factors 
assessed using non-simultaneous UV/optical data are not heavily biased.

\begin{figure}
\includegraphics[trim=20 150 30 170,clip, width=0.96\linewidth]{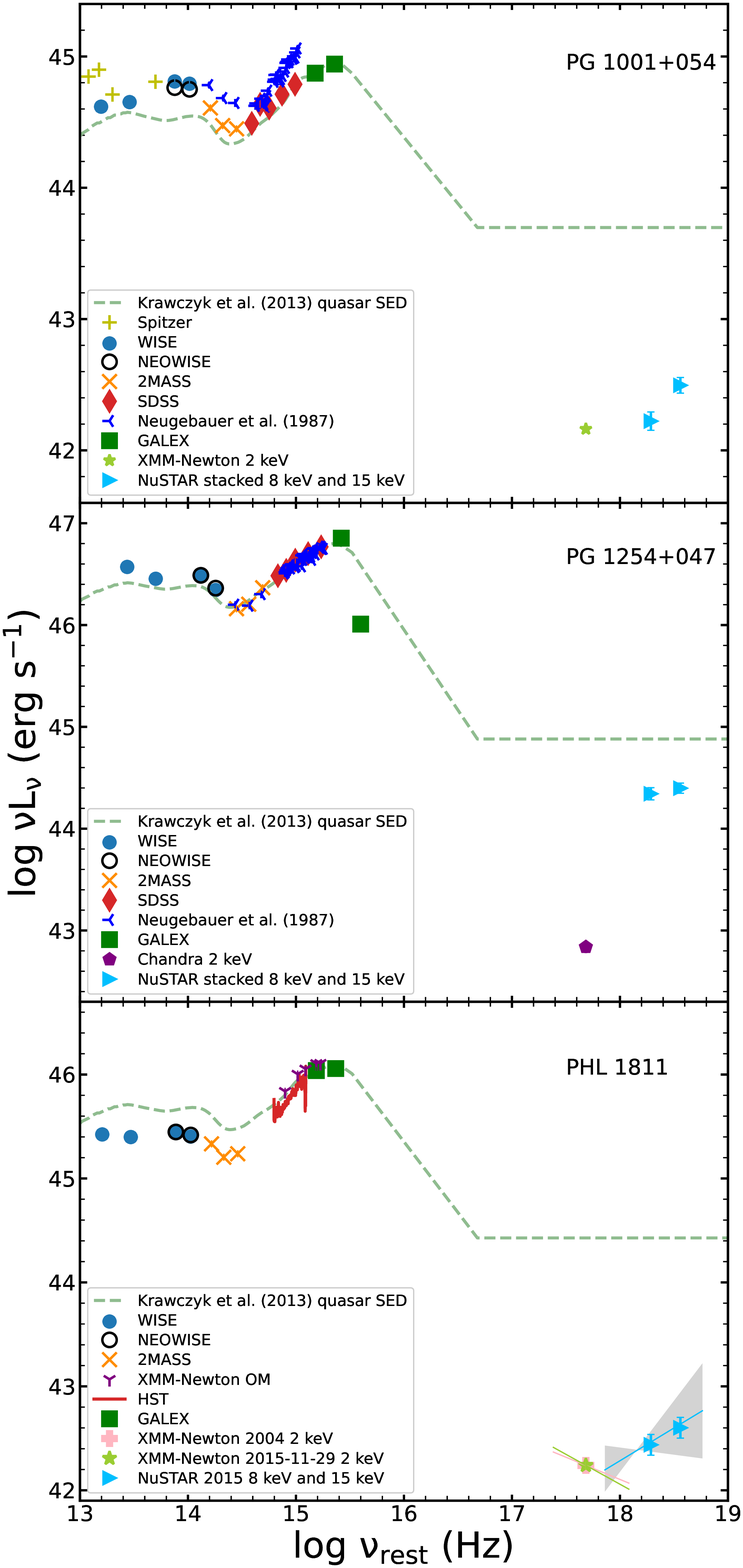}
\caption{SEDs of the three quasars (see Section 3.3 for the IR--UV data). 
The 2~keV luminosities were derived from the power-law spectral fitting 
of the \chandra\ or \xmm\ spectra (Section~3.1). The 8~keV 
and 15~keV luminosities were derived from 
the \nustar\ photometric results (Section~3.2).
For PHL~1811, we show the soft X-ray spectral slopes (with negligible uncertainties)
from the two \xmm\
observations and the hard X-ray spectral slope
(with the shaded area representing the uncertainties)
constrained from the \nustar\ observation, which differ beyond
the 1$\sigma$ level.
The green dashed curve in each panel shows the mean quasar SED 
from \citet{Krawczyk2013} normalized to the 2500~\AA\ luminosity; the X-ray 
component is a $\Gamma=2$ power-law continuum that follows
the $\alpha_{\rm OX}$--$L_{\rm 2500~{\textup{\AA}}}$ relation.
}           
\end{figure}

\section{Discussion}

The three quasars in this study were considered among the best candidates for 
intrinsically X-ray weak quasars, with coronae 
that produce weaker X-ray emission
than expected from 
the $\alpha_{\rm OX}$--$L_{\rm 2500~{\textup{\AA}}}$ relation. 
Intrinsic X-ray weakness and X-ray obscuration are not mutually exclusive.
In fact, at least for PG~1001 and PG~1254, the presence of X-ray obscuration is 
clearly indicated by
their flat spectral shapes in the soft \hbox{X-rays} (Section 3.1).
Therefore, it was proposed in \citet{Luo2014} that the \chandra, \xmm, and \nustar\ data
of PG~1001 and PG~1254 could be explained by intrinsically weak X-ray continua modified
by Compton-thin absorption.
In the subsections below, we first suggest that X-ray obscuration is also present in PHL~1811
given the hard X-ray constraints. We then propose that, as an alternative to the intrinsic X-ray weakness
$+$ X-ray obscuration scenario,  
the soft and hard X-ray weakness of these quasars can be uniformly explained
under an X-ray obscuration-only scenario,
without invoking the extra mechanism of intrinsic X-ray weakness.
The X-ray absorber in this case might be a 
clumpy accretion-disk wind that provides variable 
partial-covering absorption to the nuclear X-ray emission.

\subsection{Presence of X-ray Obscuration in PHL 1811}

PHL 1811 was considered an intrinsically X-ray weak quasar without any absorption based mainly on
the following properties: (1) significantly weak emission with a steep spectral shape
($\Gamma=2.3\pm0.1$)
in the 2004
\xmm\ \hbox{0.3--5}~keV spectrum,
and (2) flux variability by a factor of $\approx4$ in 12 days \citep{Leighly2007}.
Hard X-ray ($>5$ keV) data were not available previously.
The 2004 \xmm\ spectrum of PHL 1811 is dominated by background above 5~keV, with a 
$P_{\rm B}$ value (see Section 3.2) of only 0.1 in the 5--10 keV band.
In the deeper 2015
\xmm\ observation, PHL 1811 was significantly detected in the 5--10~keV band,
with a $P_{\rm B}$ value of $1.0\times10^{-4}$ ($3.9\sigma$), allowing investigation
of its hard X-ray properties. We first derived its photometric properties following
the same procedure described in Section~3.2, though applied to the 0.3--10~keV band instead
of the 3--24~keV band as for the \nustar\ data. The net source counts (without aperture
correction) in 
the \hbox{0.3--2~keV} and 2--10 keV bands are $872_{-34}^{+35}$ and $174^{+21}_{-20}$,
respectively, and the resulting 0.3--10 keV effective photon index 
($\Gamma_{\rm eff,0.3-10}$) is $2.0\pm0.1$ from
the band ratio of the above two bands and the spectral response files. 
If adopting a different set of energy bands,
the 0.3--5~keV and 5--10 keV bands, the derived counts are $1000^{+39}_{-38}$ 
and $47^{+14}_{-13}$, 
and
the $\Gamma_{\rm eff,0.3-10}$ value becomes $1.8^{+0.1}_{-0.2}$. 
The slightly different $\Gamma_{\rm eff,0.3-10}$ values 
suggest that the spectral shape deviates from a simple
power law.  

We then fit the 2015 \xmm\ 0.3--10 keV spectrum with a simple power-law model
using XSPEC, under the assumption that PHL~1811 is an unabsorbed, 
intrinsically \xray\ weak source. The spectrum
was grouped with at least one count per bin, and the
W statistic (Footnote \ref{wstat}) was used. The resulting photon index is 
$\Gamma=2.57\pm0.09$; if limiting the energy range to 0.3--5~keV, the resulting 
$\Gamma$ ($2.63\pm0.09$) is consistent within the errors.
The data and the best-fit model are shown in Figure~4a,
and there 
are significant residuals above
$\approx3$~keV.
It is possible that a Compton-reflection component 
from a distant reprocessor (e.g., the torus) contributes to the hard X-ray excess
emission. We thus added an XSPEC \texttt{pexrav} component to the model with its
photon index and
normalization tied to those of the power-law component; the only free parameter
is the reflection factor (reflection fraction) and the other parameters are fixed at their default values. 
The best-fit results are shown
in Figure~4b. The hard X-ray excess can be explained by the Compton-reflection component,
but an unrealistically large reflection factor ($\approx23$) is 
required; the reflection factors for AGN samples are typically $\lesssim2$ \citep[e.g.,][]{deRosa2008,Ricci2011,Ricci2017,Panagiotou2019}.
We fixed the power-law photon index to 
2.57 in the above test; when was allowed to vary, 
a larger $\Gamma$ ($\approx3.0$) and an even 
larger reflection factor ($\approx81$) 
was derived.
Therefore, the 2015 \xmm\ spectrum of PHL~1811 cannot be well described by
a simple unabsorbed power-law continuum plus a reasonable amount of Compton reflection.

{In Figure 4a, the hard X-ray residuals peak around observed-frame 
6~keV (rest-frame $\approx 7$~keV), and we thus also considered a model where
the hard X-ray excess has some contribution from a broad Fe K$\alpha$ line
produced via relativistic disk reflection
\citep[e.g.,][]{Ross2005,Fabian2013}.
We fit the 2015 \xmm\ 0.3--10 keV spectrum with the XSPEC
\texttt{relxill} model \citep[][]{Dauser2014,Garcia2014}.
The free parameters are $\Gamma$, SMBH spin, ionization parameter,
inclination angle, and reflection factor (reflection fraction), and 
the other parameters were fixed at their default values.
This model describes well the spectrum with similar residuals to 
those in Figure 4b, with a best-fit $\Gamma$ value of $2.02\pm0.06$
and a best-fit reflection factor of $10^{+65}_{-1}$. 
The $\Gamma$ value is smaller than 2.57 because the soft X-rays are now
dominated by
ionized-disk reflection instead of an intrinsic power-law continuum.
The 
reflection factor is still large, but unlike the Figure~4b modeling with
a distant
reflector, relativistic reflection from the inner accretion disk could
produce an extremely large reflection factor 
if much of the coronal emission cannot 
reach the observer due to the light bending effects near the SMBH
\citep[e.g.,][]{Dauser2014}. Nevertheless, a reflection-dominated 
spectrum with strong light bending effects 
is still in contrast with the scenario of intrinsic X-ray weakness
where the observed spectrum should be dominated by the intrinsic power-law 
continuum from a weak corona.}

\begin{figure*}
\includegraphics[trim=10 10 10 20,clip, width=0.49\linewidth]{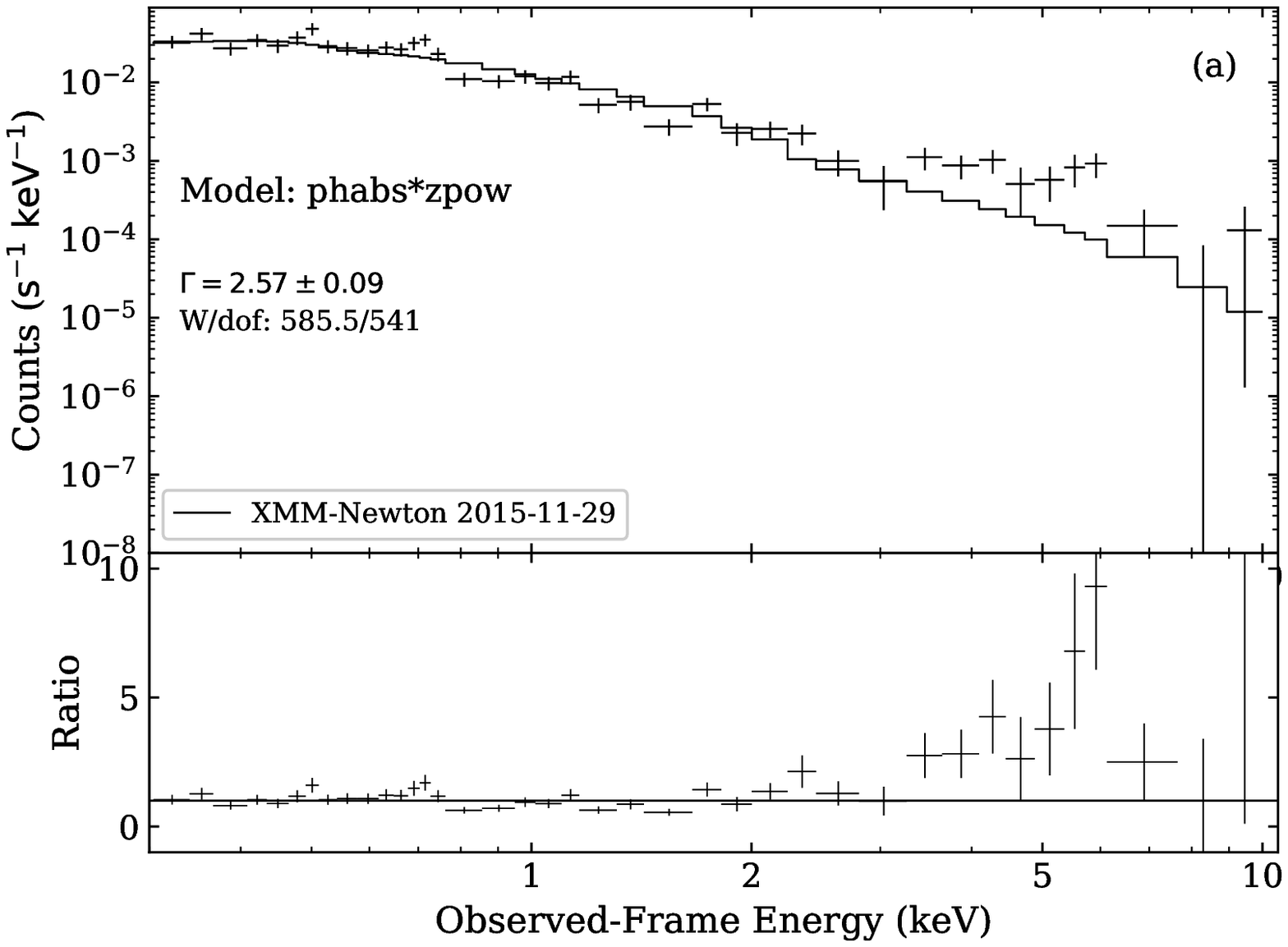}
\includegraphics[trim=10 10 10 20,clip, width=0.49\linewidth]{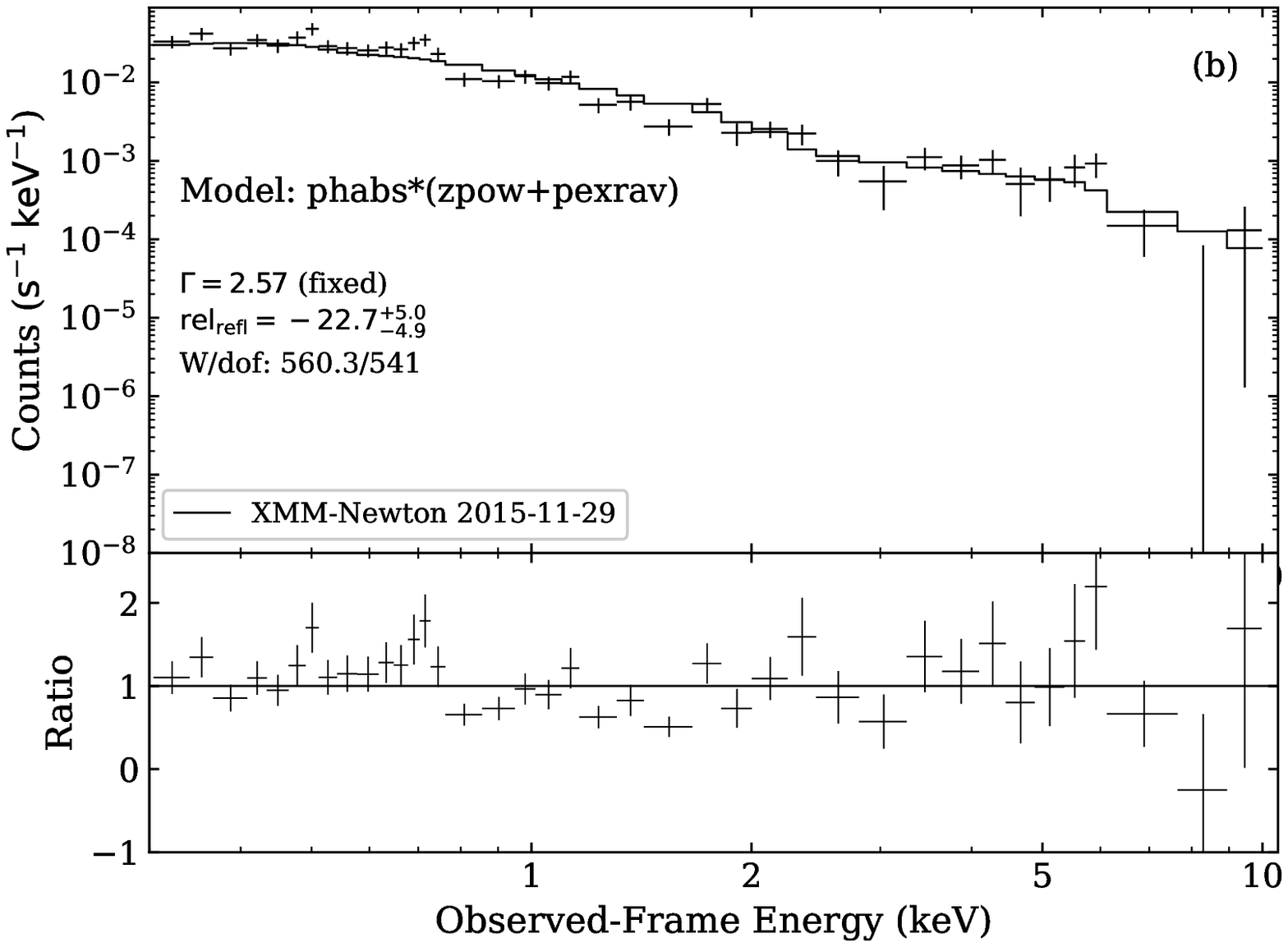}
\caption{The 2015 \xmm\ spectrum for PHL~1811
overlaid with the best-fit (a) simple power-law model and (b)
simple power-law plus Compton-reflection model (with a fixed photon index).
The spectra are
grouped for display purposes only.
The bottom panels display the ratios of the data to the best-fit models.
Compared to the simple power-law model,
there are significant fitting residuals above $\approx3$~keV, which
require an unrealistically large reflection factor ($\approx23$) to explain.
The spectrum thus cannot be well described by
a simple unabsorbed power-law continuum plus a reasonable amount of Compton reflection.
}
\end{figure*}

\begin{figure*}
\includegraphics[trim=10 10 10 20,clip, width=0.49\linewidth]{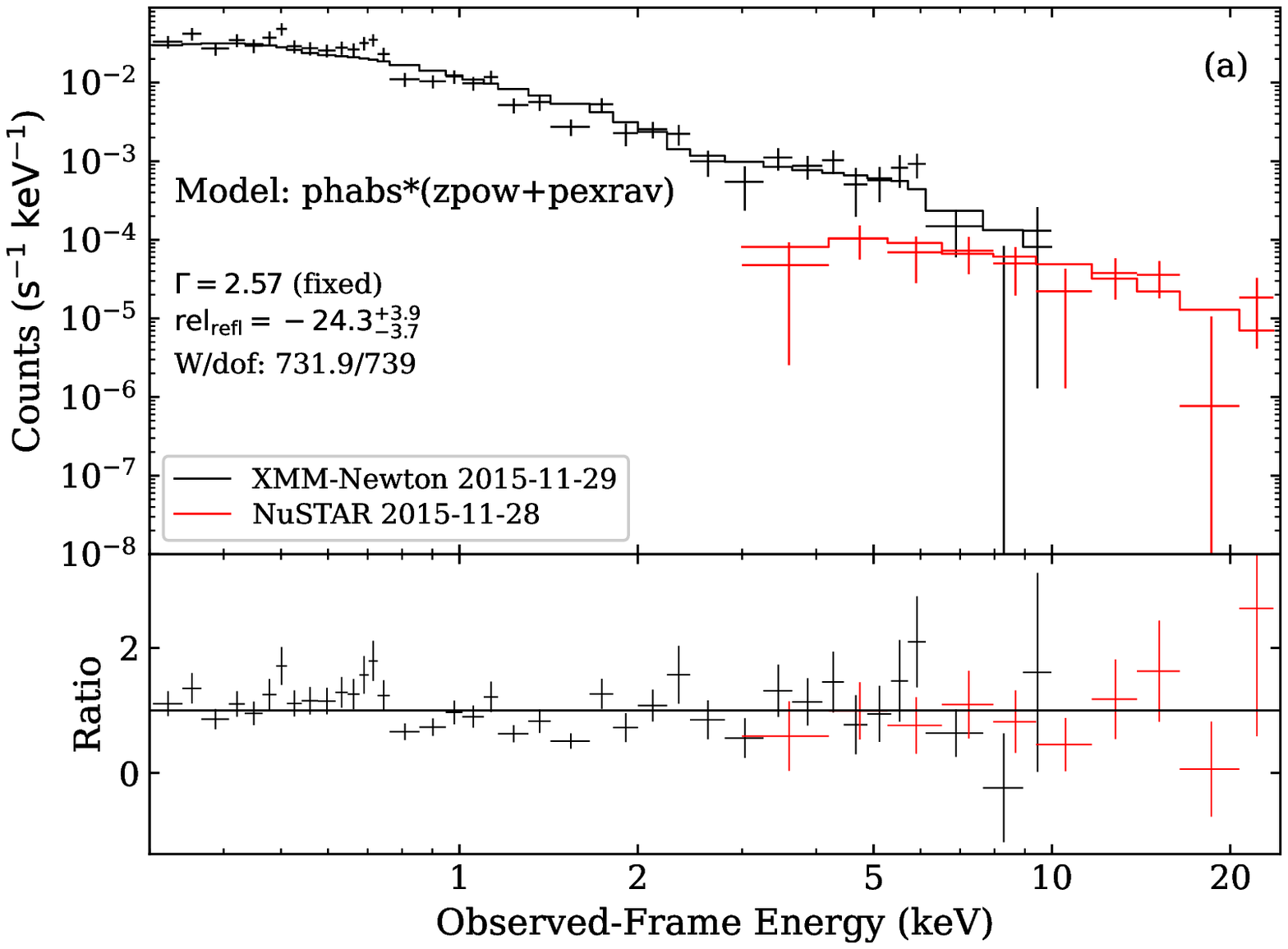}
\includegraphics[trim=10 10 10 20,clip, width=0.49\linewidth]{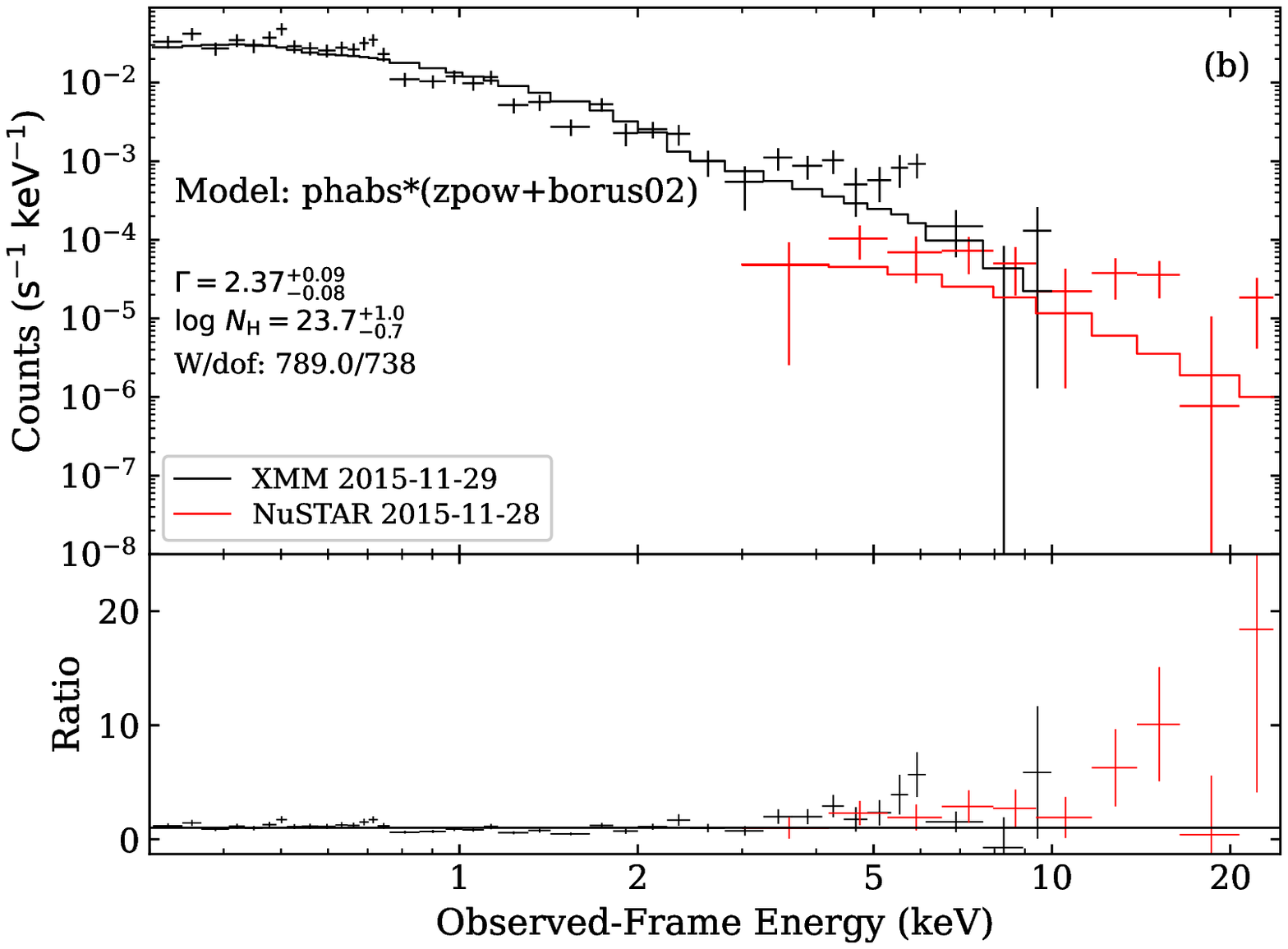}
\caption{
Similar to Figure 4, displaying the 2015 \xmm\ and \nustar\ spectra for PHL~1811
overlaid with the best-fit power-law plus Compton-reflection model.
The Compton-reflection component was modeled with (a) \texttt{pexrav}
or (b) \texttt{borus02}. An unrealistically large reflection
factor ($\approx24$) is required in the \texttt{pexrav} model, while the
self-consistent \texttt{borus02} Compton reflection is not sufficient to
account for
the excess emission in the hard X-rays.}
\end{figure*}

The 3--24 keV spectral shape constrained from the 2015 simultaneous \nustar\
observation, $\Gamma_{\rm eff}=1.4^{+0.8}_{-0.7}$
(Section~3.2), provides additional support that the spectral shape likely deviates from 
a steep unabsorbed power law.
We thus applied the above \texttt{zpow} $+$ \texttt{pexrav} model to jointly
fit the 2015 \xmm\ $+$ \nustar\ spectra and investigate if the spectra 
can be explained by an unabsorbed, intrinsically
weak power-law continuum plus typical Compton reflection.
The results are shown in Figure~5a, and they are consistent with those for the 
\xmm\ spectrum alone (Figure~4b), requiring a huge reflection factor.
We then tested replacing the \texttt{pexrav} component with 
the self-consistent Compton-reflection model \texttt{borus02} \citep{Balokovi2018} 
with the photon index and normalization tied to those of the power-law component.
The complete XSPEC model is
\texttt{phabs} $*$ (\texttt{zpow} $+$ \texttt{atable\{borus02.fits\}}),
and the free parameters are the power-law normalization, 
photon index ($\Gamma$), and column density
($N_{\rm H}$) of the reprocessed component.\footnote{
The other parameters of the \texttt{borus02} model 
are fixed at the default values including a 
high-energy cutoff of 300~keV, an inclination angle of 60\degr, a
torus covering factor of 0.5 (corresponding to a half-opening angle of 60\degr), 
and an iron relative abundance of 1.
\label{footnotebo}} 
The best-fit results
are displayed in Figure~5b. There are still significant fitting 
residuals above $\approx5$~keV,
demonstrating again that a typical level of Compton reflection
cannot account for 
the excess emission in the hard X-rays. 

The 2015 \xmm\ and \nustar\ observations of PHL~1811 thus reveal that, 
in addition to a steep power-law ($\Gamma=2.63\pm0.09$) continuum in the 0.3--5~keV
band, there is significant hard X-ray ($\gtrsim5$~keV) excess emission.
The excess emission can be modeled with Compton reflection of a 
soft X-ray continuum that is much stronger (by a factor of $\gtrsim20$) 
than the observed one. We note that, from the best-fit
\texttt{zpow} $+$ \texttt{pexrav} results in Figure~5a, PHL~1811 was intrinsically
X-ray weak by a factor
of $\approx140$ compared to the \citet{steffen2006}
$\alpha_{\rm OX}$--$L_{\rm 2500~{\textup{\AA}}}$ relation.
If we instead assume that PHL~1811 was intrinsically X-ray normal (i.e., raising
the power-law normalization by a factor of 140) and add a scaling factor to 
the power-law component \texttt{c}\,*\,\texttt{zpow} $+$ \texttt{pexrav}, 
the spectra can be equally well described with $\texttt{c}=0.7\%$ ($1/140$) and a 
reasonable reflection factor of 0.17 ($24/140$). 
A physical interpretation of the above model is that PHL~1811 was affected by
Compton-thick absorption and the observed \xmm\ and \nustar\ spectra were dominated
by a
fraction
of the intrinsic continuum scattered by a large-scale highly ionized ``mirror''
($f_{\rm scatter}$; typically within a few percent; e.g., \citealt{Turner1997}; \citealt{Cappi2006}; \citealt{Ueda2007}; \citealt{Winter2009}; 
\citealt{Yamada2020}; \citealt{Gupta2021}) in the soft X-rays and a reprocessed component from the absorber
in the hard \hbox{X-rays}. The steep soft continuum could also represent a 
fraction ($f_{\rm leak}$) of the intrinsic continuum leaking through a clumpy absorber. 
Therefore, instead of employing intrinsic X-ray weakness plus an unrealistically 
large reflection factor, we could interpret the 2015 \xmm\ and \nustar\ data
of PHL~1811
with Compton-thick obscuration.

In addition, the second property mentioned above, the short-term soft X-ray variability
between the two \chandra\ observations in 2001 (e.g., see Figure~1), 
was used to argue against
a large-scale scattered component ($f_{\rm scatter}$ above) 
dominating the \hbox{0.3--5 keV} X-ray spectrum. 
However, from recent investigations
of extreme X-ray weakness and extreme X-ray variability among super-Eddington 
accreting AGNs (e.g.,
SDSS J$075101.42+291419.1$ and samples from \citealt{Liu2019,Liu2021}; SDSS J$081456.10+532533.5$ from Huang~J. et al. in prep.), it appears plausible
that such weak, steep, and variable soft \xray\ emission 
might originate from variable fractions ($f_{\rm leak}$) of leaked
intrinsic continuum through 
a large solid-angle, high column-density, clumpy (partial-covering) absorber.
Thus fast 
variability and
steep spectral shapes in the soft X-rays do not
necessarily rule out X-ray obscuration.
In summary, unlike the model considered previously where PHL~1811 lacks 
absorption, hard \hbox{X-ray} data suggest that \xray\ obscuration
may well be present.

\subsection{An Obscuration-Only Scenario Without Intrinsic X-ray Weakness}

The 2013 \nustar\ observations of PG~1001 and PG~1254, 
with no hard-band detections or $\Gamma_{\rm eff}$ measurements,
were not sufficiently constraining
to establish that intrinsic X-ray weakness must be present, which
motivated the present study with deeper \nustar\ observations.
The deeper \nustar\ observations now provide hard-band detections and
$\Gamma_{\rm eff}$ measurements (albeit with large uncertainties). 
The PG~1001 spectral shape
appeared flatter in the 2020 observation
($\Gamma_{\rm eff}=0.4^{+0.6}_{-0.9}$ compared to 
$\Gamma_{\rm eff}>1.5$), 
suggesting the presence of absorption and probably spectral-shape evolution
between these two epochs.
The stacked $\Gamma_{\rm eff}$ value ($1.0^{+0.5}_{-0.6}$) from the two \nustar\ observations is still 
small compared to the typical value of $\approx2$ for an unabsorbed spectrum.
The PG~1254 spectral slope appears typical for unobscured quasars
($\Gamma_{\rm eff}=1.8^{+0.5}_{-0.4}$ in the 2019 observation), 
but it is also clear
that this quasar is X-ray weak by only a factor of a few in the hard X-rays
(see Figure 3). 
Considering its significant soft X-ray weakness and the flat spectral shape 
in the soft X-rays (Section 3.1),
the nominal hard X-ray spectral
shape in the \nustar\ data might be explained by Compton-thin absorption.
Since PG 1254 is at $z=1.026$,
we are likely observing the
penetrating hard X-rays with \nustar\ through an absorber with
a large but Compton-thin $N_{\rm H}$ value;
this would explain the strong
X-ray weakness at 2~keV and much reduced \xray\ weakness at higher energies. 
As discussed in Section 4.1 above, hard X-ray data suggest 
that \hbox{X-ray} obscuration is also
present in PHL~1811.

Motivated by these \nustar\ results, we argue that intrinsic X-ray weakness
is probably not required to explain the extreme X-ray weakness of these quasars. 
We explore the possibility of interpreting universally
the \nustar, \chandra, and \xmm\ spectra with an 
obscuration-only scenario
where these quasars are intrinsically \hbox{X-ray} normal (following
the $\alpha_{\rm OX}$--$L_{\rm 2500~{\textup{\AA}}}$ relation). 
We investigate below, via XSPEC spectral fitting, if the multi-epoch soft and hard 
X-ray spectra can be described by
nominal-strength X-ray emission modified by
our adopted obscuration model with the 
absorber parameters (the column density and partial-covering fraction) allowed
to vary.
Since the X-ray data do not have sufficient statistics for complex spectral fitting, 
we had to simplify the model and fix many of the model parameters.
Moreover, we actually cannot rule out the scenario of obscuration $+$ intrinsic X-ray weakness,
which has an extra degree of freedom (i.e., normalization of the \xray\
continuum) compared to the obscuration-only scenario.
Our focus here is to investigate if we can explain the observed X-ray emission without
involving 
intrinsic X-ray weakness from, e.g., an anomalous corona.

Although the \texttt{pexrav} model appears able to describe the PHL~1811
spectra well (Section 4.1 and Figure~5a),
it does not provide constraints
on the absorption column densities. We thus employed the 
self-consistent \texttt{borus02} model 
to describe the reprocessed component 
from the absorber.
The XSPEC spectral model is 
\begin{eqnarray*}
\texttt{phabs}\,*\,\left(\texttt{zphabs}\,*\,\texttt{cabs}\,*\,\texttt{c}_1\,*\,\texttt{zpow}\right.\\
              +\,\left.\,\texttt{c}_2\,*\,\texttt{zpow}+\,\texttt{atable\{borus02.fits\}}\,\right).
\end{eqnarray*}
In this model, \texttt{phabs} accounts for the Galactic absorption,
and \texttt{zpow} is the intrinsic
power-law continuum that is X-ray normal with respect to the 
\citet{steffen2006}
$\alpha_{\rm OX}$--$L_{\rm 2500~{\textup{\AA}}}$ relation.
A large fraction (\texttt{c}$_1$) of 
the intrinsic continuum is modified by heavy neutral absorption (\texttt{zphabs}) 
and Compton scattering (\texttt{cabs}).
The absorber is clumpy, allowing a
fraction ($f_{\rm leak}=$\texttt{c}$_2=1-$\texttt{c}$_1$) 
of the intrinsic continuum to leak through. There is probably also a
large-scale scattered component ($f_{\rm scatter}$).
Since the leaked component usually dominates, we treated them together and do not 
separate $f_{\rm scatter}$ from 
$f_{\rm leak}$ in 
the following study.
The reprocessed component from the absorber is modeled with
\texttt{borus02} with the normalization and photon index tied
to those of \texttt{zpow}.
We fixed the inclination angle to 60\degr\ and allowed
the absorber covering factor to vary.
The other \texttt{borus02} parameters were fixed at the default values (Footnote \ref{footnotebo}).
We also tied the absorption column densities ($N_{\rm H}$)
in the \texttt{zphabs}, \texttt{cabs}, and \texttt{borus02} components.
The emergent spectrum is thus a combination of the transmitted (absorbed) component, 
leaked component, and
reprocessed component. A schematic illustration of the setup is shown in
Figure~6a, and an example of the X-ray spectral components for a given set of the
absorber parameters is displayed in Figure~6b. 

{
There are a couple of caveats regarding the above model. 
First, the absorber
is probably partially ionized instead of being neutral. 
Ionized absorption produces distinctive spectral features below 
$\approx1$~keV, but the  
continuum shapes above
$\approx1$~keV are similar to those from neutral absorption
unless the ionization level is high
(e.g., Figure~1 of \citealt{Netzer1993}). For the three quasars studied here,
PG~1254 has few photons below 1~keV, PHL~1811 shows no absorption signatures
below $\approx5$~keV, and PG~1001 has clear soft X-ray excess emission 
in the 0.3--1~keV band which was interpreted with ionized absorption
\citep{Schartel2005}. Their $>1$~keV spectra do not have 
sufficient photon statistics to distinguish ionized absorption from
neutral absorption or constrain reliably ionization parameters.
Therefore, in the above model, we adopted neutral absorption for
simplicity, and we did not use the $<1$~keV \xmm\ or \chandra\ data
for PG~1001. The soft excess of PG~1001 in the obscuration scenario will 
be discussed in Section~4.2.1 below.
Second, since there is no optical/UV extinction, the absorber
should not be the torus described in the \texttt{borus02} model with a
toroidal geometry. Instead, it is likely a small-scale clumpy dust-free wind
launched from the accretion disk (see Section~4.3 below).
Therefore, the
\texttt{borus02} model does not provide an accurate description of the
reprocessed emission (both the continuum and the Fe K$\alpha$ line) from
the absorber (e.g., wind). However, since 
our purpose here is not to recover precise absorber parameters 
but to simply investigate if the
obscuration-only scenario is a valid alternative to the scenario of 
intrinsic X-ray weakness
$+$ obscuration,
and
the current simplified model appears 
able to explain reasonably well the multi-epoch
\hbox{X-ray} spectra of the three quasars (as discussed below), 
we defer detailed modeling to future studies which
likely will require much better
spectral quality.}


\begin{figure*}
\includegraphics[trim=65 65 20 20,clip, width=0.55\linewidth]{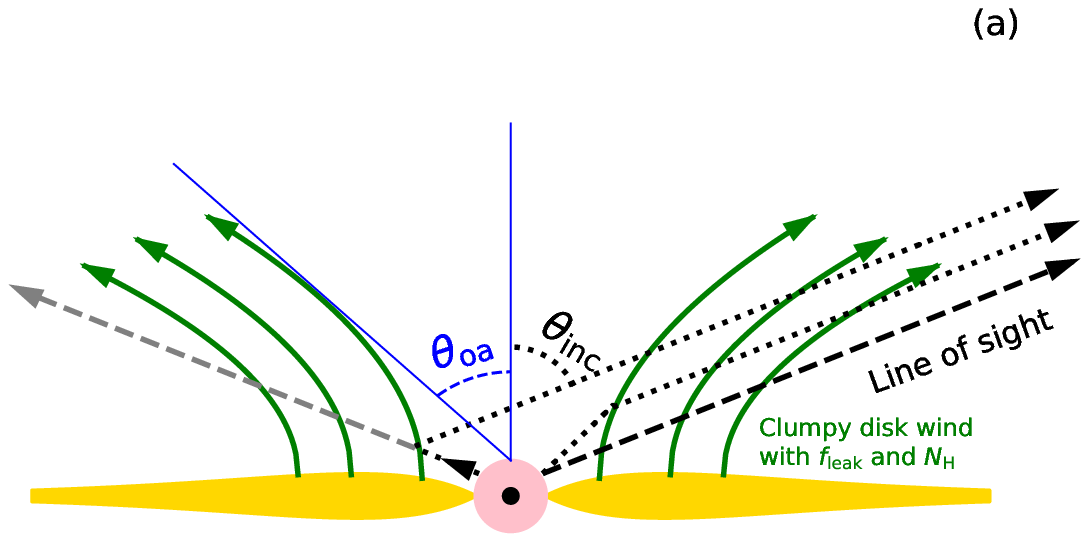}
\includegraphics[trim=10 10 10 20,clip, width=0.45\linewidth]{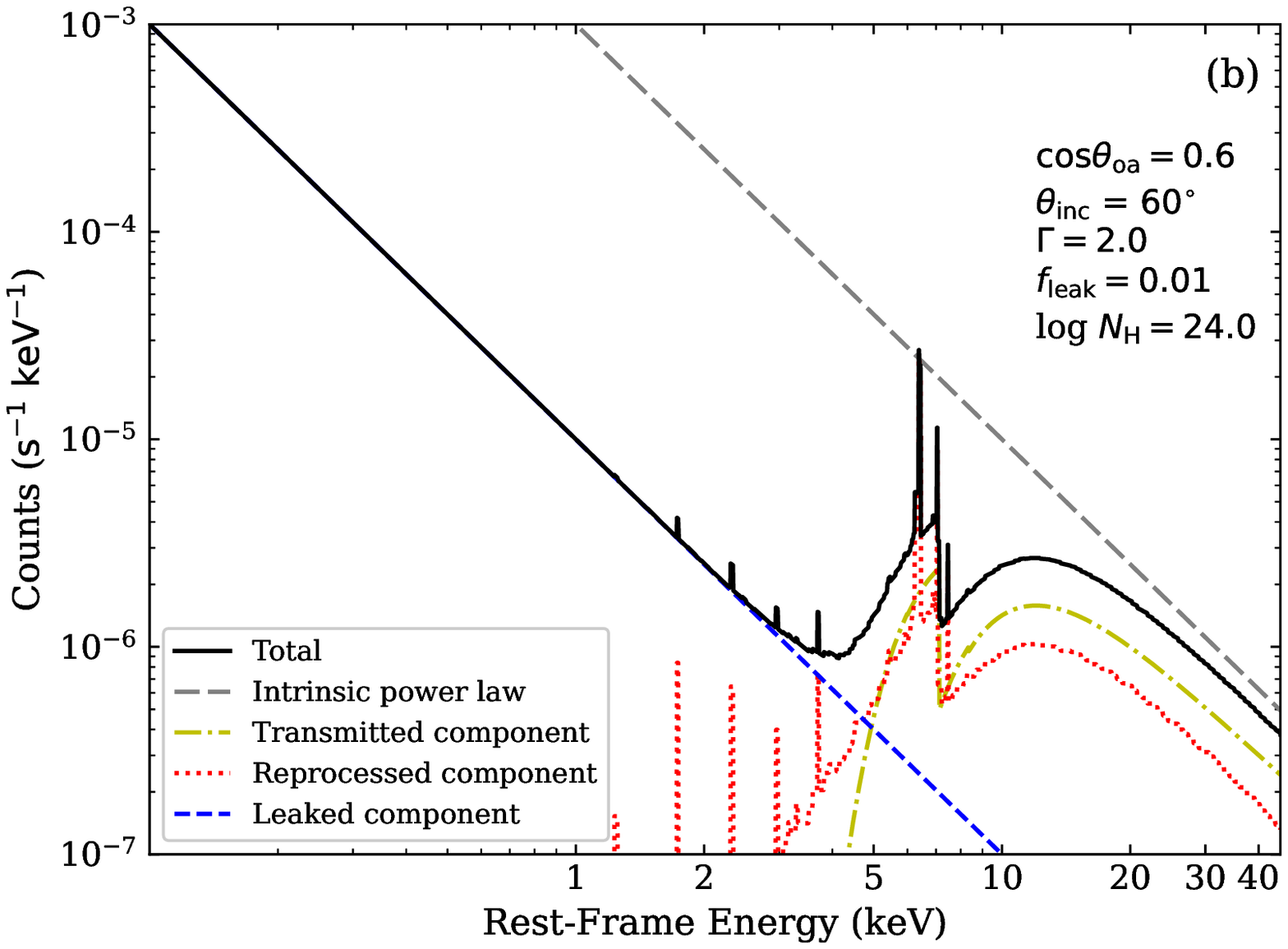}
\caption{(a): A schematic diagram of the obscuration scenario. The black dot
represents the SMBH, surrounded by the \hbox{X-ray} corona shown
in pink. The yellow region represents the accretion disk.
The green arrows represent the absorber
(e.g., an outflowing disk wind) with a half-opening angle
of $\theta_{\mbox{\scriptsize oa}}$ and a covering factor of $\cos\theta_{\mbox{\scriptsize oa}}$. 
The system inclination angle is
$\theta_{\mbox{\scriptsize inc}}$. The dashed line traces the transmitted (absorbed)
radiation through the absorber in the direction of the observer (transmitted component).
The dotted curves trace the radiation reflected from the opposite
side of the absorber and the radiation
scattered to the line of sight through the absorber (reprocessed component).
The absorber is clumpy allowing a
fraction ($f_{\rm leak}$) of the intrinsic continuum to leak through without being absorbed 
(leaked component).
There is probably also a fraction ($f_{\rm scatter}$) of the intrinsic
continuum scattered to the line of sight by a large-scale highly 
ionized ``mirror'' (not shown in the cartoon). We treated $f_{\rm leak}$
and $f_{\rm scatter}$ together in this model.
{(b): An example of the X-ray spectral components for a given set of the
absorber parameters. The black solid curve shows the total observed model spectrum, while
the gray long-dashed line, yellow dashed-dotted curve, red dotted curve, 
and blue dashed line represent 
the intrinsic power law, transmitted component, reprocessed component, 
and leaked component, respectively.}}
\end{figure*}

We applied the above model to explain the multi-epoch
\hbox{X-ray} spectra of the three quasars.
We jointly fit the \nustar, \chandra, and \xmm\ spectra 
for each of the three quasars.
The free parameters are $\Gamma$,
$f_{\rm leak}$ (\texttt{c}$_2$), \nh, and the absorber
covering factor ($\cos\theta_{\mbox{\scriptsize oa}}$ in
Figure~6a); $f_{\rm leak}$ and $N_{\rm H}$ were allowed to vary between
the observations while the other two parameters were tied.
In addition, 
for the latest \nustar\ observation of PG~1001 and the two
\nustar\ observations of PG~1254, we tied
the $f_{\rm leak}$ parameter to that of the non-simultaneous
\chandra\ observation,
as the \nustar\ spectra are not sensitive
to this parameter (high-energy spectra do not have a significant leaked component).
These quasars are considered to be intrinsically \hbox{X-ray}
normal, with the intrinsic $f_{\rm 2keV}$ values fixed at those expected from the \citet{steffen2006}
$\alpha_{\rm OX}$--$L_{\rm 2500~{\textup{\AA}}}$ relation. 
The $f_{\rm 2keV}$ and $\Gamma$ values define the 
intrinsic continua (normalizations of \texttt{zpow}).
For each object, the \texttt{zpow} normalization was first fixed 
at the value derived with $\Gamma=2$, and then an iterative procedure was performed to
adjust the \texttt{zpow} normalization according to the best-fit $\Gamma$ value.
A few
iterations were needed until the intrinsic continuum converged.

The best-fit parameters are listed in Table~3 
and the best-fit models are displayed in Figure~7.
We also list in Table~3 the main components (transmitted, reprocessed, and/or leaked) 
that dominate the emergent spectrum of each observation.
The $\Gamma$ values for PG~1001 and PHL~1811 pegged at 2.6, the 
upper bound allowed by the \texttt{borus02} model.
Several 
$N_{\rm H}$ values do not have upper errors as the emergent spectrum is
dominated by the leaked component which is not sensitive to $N_{\rm H}$;
a few of these pegged at $\log N_{\rm H}/{\rm cm}^{-2}=25.5$, the 
upper bound allowed by the \texttt{borus02} model.
The fitting results are acceptable overall 
considering the fitting statistics ($W/{\rm dof}$; Figure~7)
and residuals, indicating that the obscuration scenario is able to 
explain the multi-epoch X-ray data without involving intrinsic X-ray weakness.
From the best-fit models, we computed the $f_{\rm w}$ values for the \nustar\ spectra, and these
are also listed in Table~3.
These hard X-ray weakness factors are comparable to those listed in Table~2,
and any discrepancy is mainly due to the different models used (a simple power-law model
is assumed in deriving the photometric results in Table 2).

\begin{deluxetable*}{lcccccccc}
\tablecaption{Best-fit Parameters for the Multi-epoch X-ray Spectral Fitting}
\tablehead{
\colhead{Object Name}                   &
\colhead{Observatory}&
\colhead{Obs. Date}&
\colhead{$\cos\theta_{\mbox{\scriptsize oa}}$} &
\multicolumn{3}{c}{Partial-Covering Absorption} &
\colhead{$f_{\rm w}^{a}$}&
\colhead{Main~Com.$^{b}$}\\
\cline{5-7}
\colhead{}&
\colhead{}&
\colhead{}&
\colhead{}&
\colhead{$f_{\rm leak}$}&
\colhead{$\Gamma$}&
\colhead{$\log N_{\rm H}$ (cm$^{-2})$}&
\colhead{}&
\colhead{}
}
\startdata
PG~1001&\xmm\     &2003 May 4&$0.54^{+0.06}_{-0.07}$&$1.9^{+0.4}_{-0.3}\times10^{-2}$&     $2.6_{-0.02}$$^d$&$23.5\pm0.1$  & --&    leak,~tra\\
PG~1001&\chandra\   &2010 Jan 11&--&$7.5^{+4.6}_{-3.5}\times10^{-3}$&     --&$23.7^{+1.3}_{-0.2}$&  --& leak,~tra\\
PG~1001&\nustar\    &2013 Jun 28&--&$0.10\pm0.05$& --&$24.3^{+0.9}_{-0.2}$& 16.7&   leak,~rep\\
PG~1001&\nustar\   &2020 May 23&--&$7.5\times10^{-3}$ (tied)&    --&$24.2\pm0.1$&52.7&        tra,~rep\\
\\
PG~1254&\chandra\    &2000 May 29 &$0.38^{+0.06}_{-0.19}$&$7.4^{+2.7}_{-2.1}\times10^{-3}$&     $2.12^{+0.03}_{-0.09}$&$25.5_{-1.0}$& --&   leak,~rep \\
PG~1254&\nustar\    &2013 Jun 8&--&-- (tied)&     --&$24.0\pm0.2$& 8.7&    tra,~rep\\
PG~1254&\nustar\    &2019 Jun 8&--&-- (tied)&     --&$23.6^{+0.2}_{-0.1}$& 3.2&    tra \\
\\
PHL~1811&\chandra\    &2001 Dec 5&$0.58^{+0.01}_{-0.02}$&$6.0^{+0.8}_{-0.7}\times10^{-3}$&     $2.6_{-0.05}$&$24.1^{+0.5}_{-0.2}$& --&  leak,~rep\\
PHL~1811&\chandra\    &2001 Dec 17&--&$(2.8\pm0.2)\times10^{-2}$&     --&$24.9_{-0.5}$& --& leak\\
PHL~1811&\xmm\    &2004 Nov 1&--&$(6.7\pm0.4)\times10^{-3}$&     --&$25.5_{-0.5}$& --&  leak\\
PHL~1811&\xmm\   &2015 Nov 29&--&$(7.4\pm0.3)\times10^{-3}$&     --&$24.6_{-0.2}$& --& leak,~rep\\
PHL~1811&\nustar\    &2015 Nov 28&--&$(1.1\pm0.8)\times10^{-2}$&     --&$24.8\pm0.2$& 98.3& leak,~rep\\
PHL~1811&\xmm\ + \nustar$^{c}$ & 2015 Nov&--& $(7.4\pm0.3)\times10^{-3}$& --& $24.8^{+0.4}_{-0.2}$&--&    leak,~rep
\enddata
\label{tbl-obs}
\tablenotetext{a}{The factor of X-ray weakness at rest-frame 8 keV derived from the best-fit model, 
for comparison with the results in Table 2.}
\tablenotetext{b}{The ``Main~Com.'' column lists the dominant component/components in the emergent spectrum: ``leak'' represents the
leaked/scattered component, ``tra'' represents the transmitted (absorbed) component through
the absorber (dashed line in Figure 6), 
and ``rep'' represents the reprocessed component from the absorber (dotted curves in Figure~6).} 
\tablenotetext{c}{In this case, the fitting parameters for the \xmm\ and \nustar\ observations were tied.}
\tablenotetext{d}{A value without an upper error is effectively bound by the 
allowed upper bound of the \texttt{borus02} model.}
\end{deluxetable*}

\begin{figure*}
\includegraphics[trim=0 10 20 20,clip, width=0.49\linewidth]{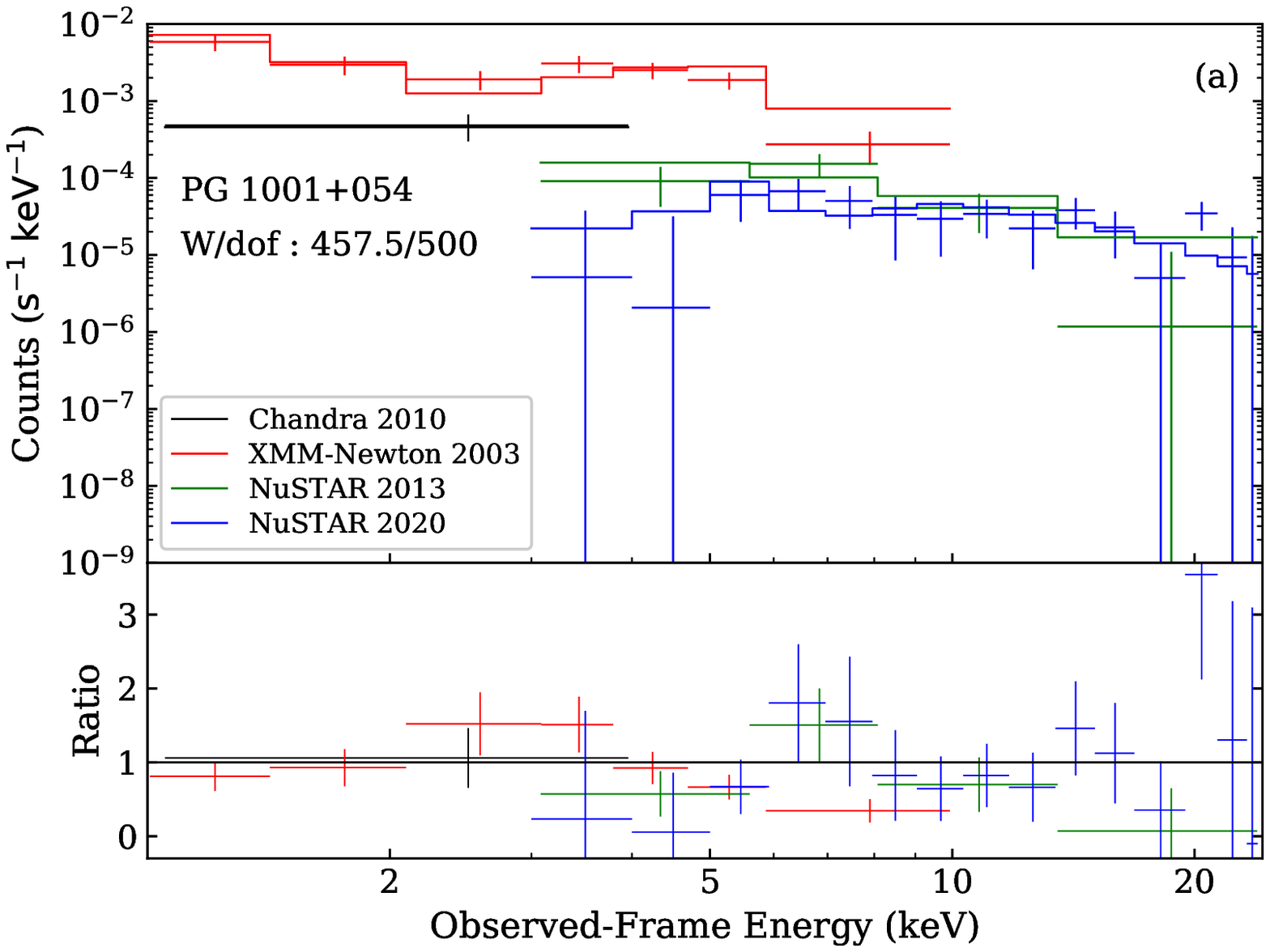} 
\includegraphics[trim=0 10 20 20,clip, width=0.49\linewidth]{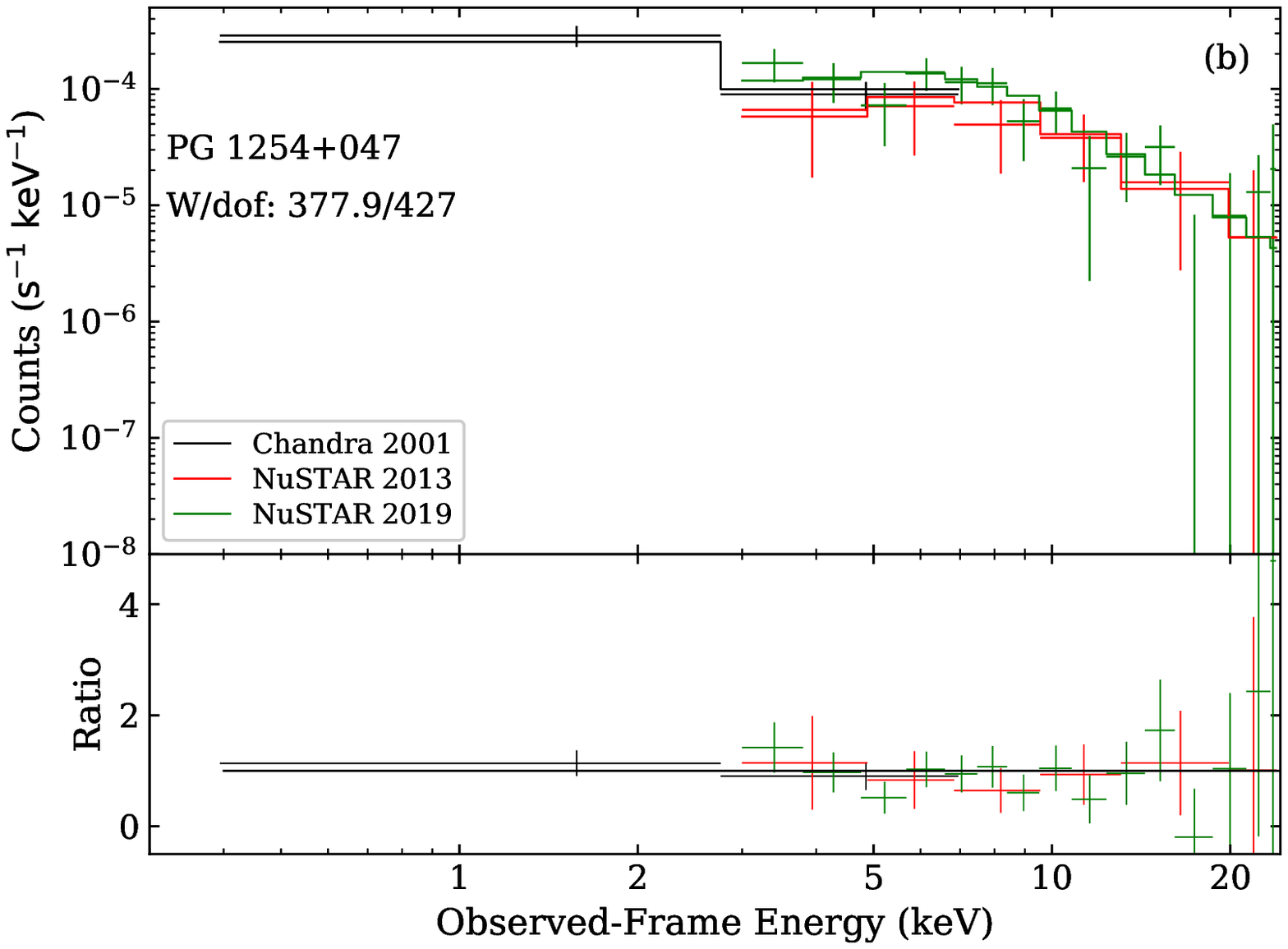}  
\includegraphics[trim=0 10 20 20,clip, width=0.49\linewidth]{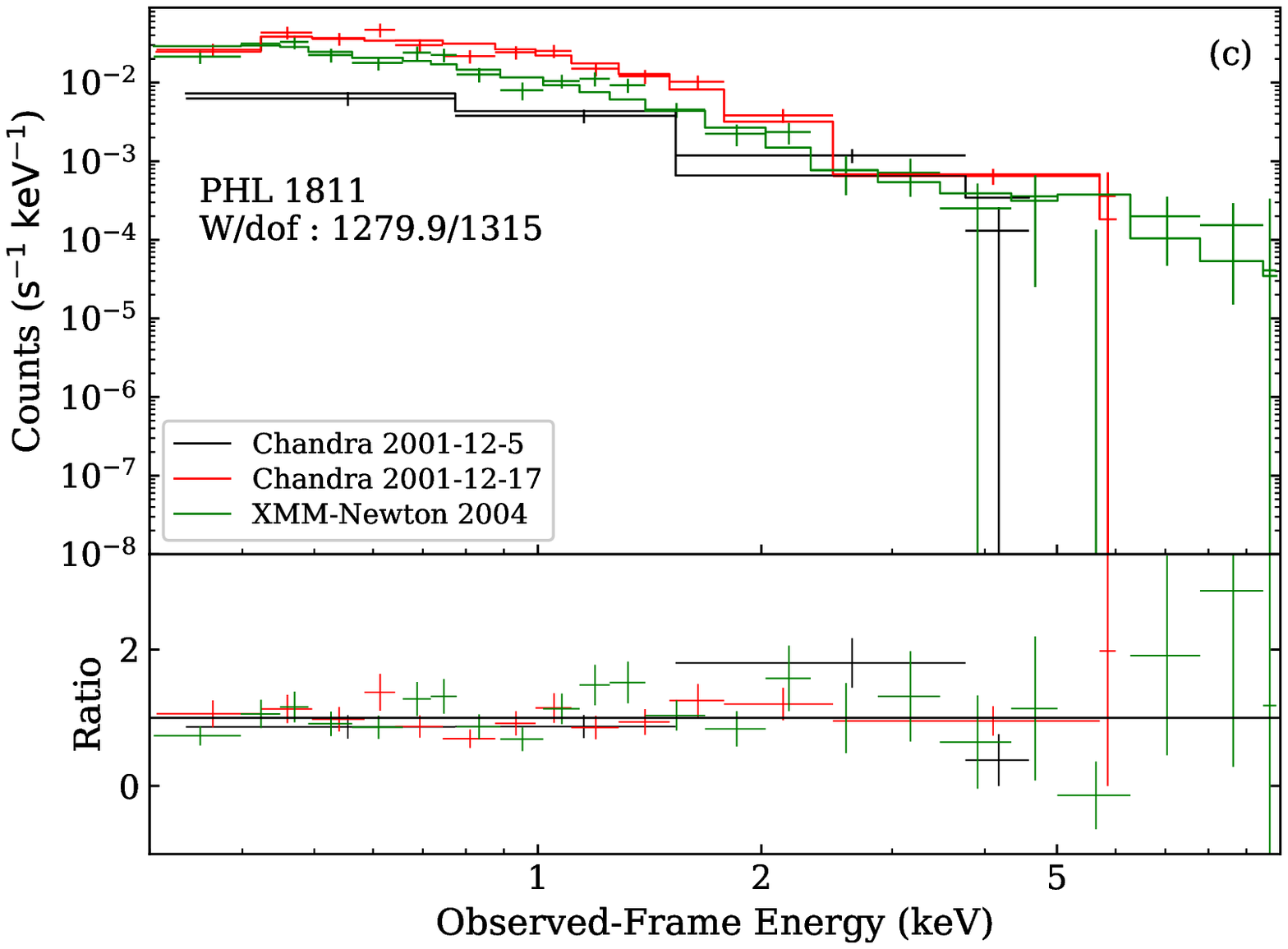}
\includegraphics[trim=0 10 20 20,clip, width=0.49\linewidth]{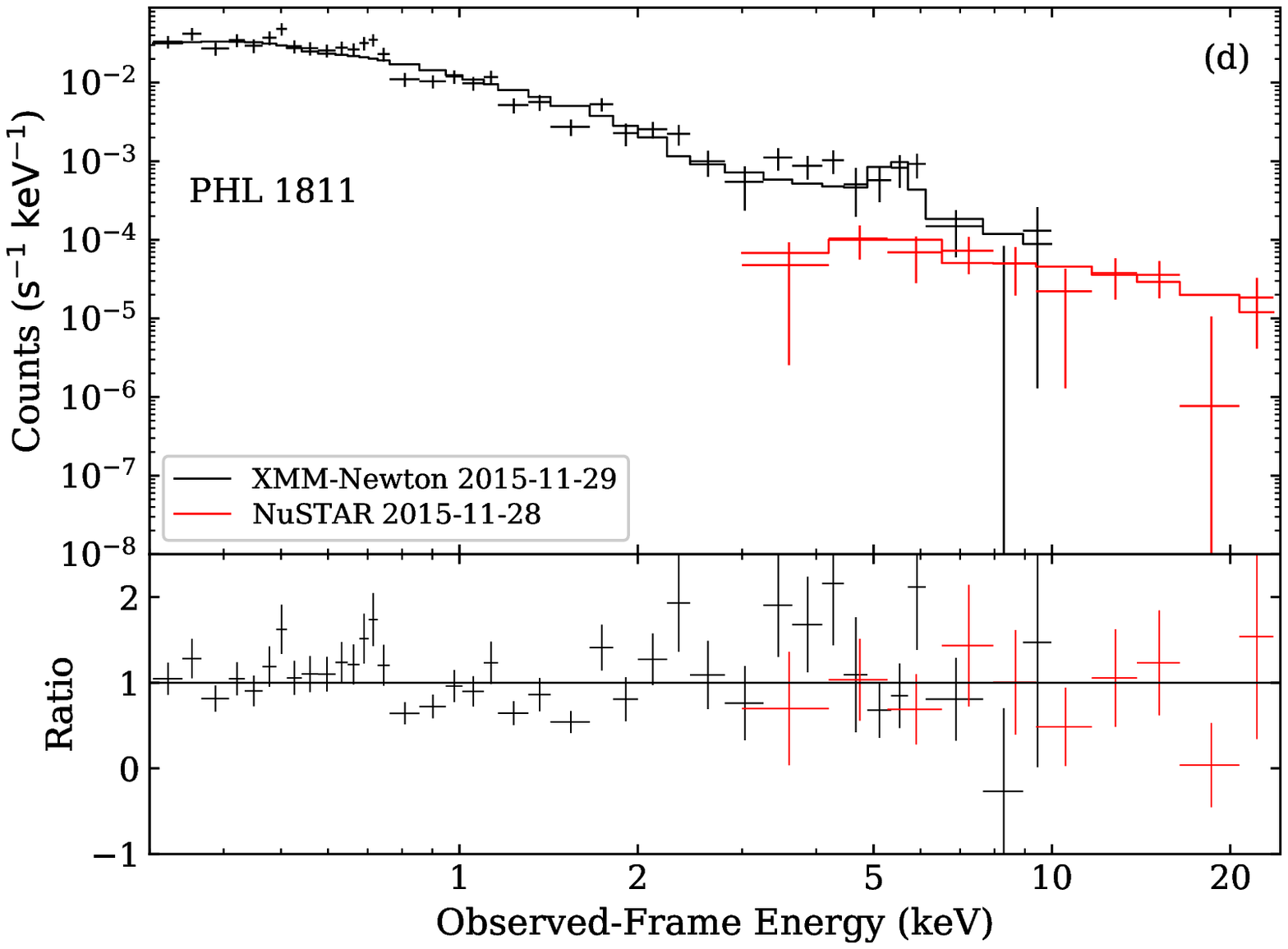}  
\caption{The \chandra, \xmm, and \nustar\ spectra for the three quasars, overlaid with
the best-fit models using the obscuration model described in 
Section 4.2. The spectra are
grouped for display purposes only.
The bottom panels display the ratios of the data to the best-fit models.
We break the five observations of PHL 1811 into two panels for
clarity. For PG~1001, the energy ranges used are 1--8~keV for the
\chandra\ observation and 1--10~keV 
for the \xmm\ observation 
due to the apparent soft X-ray excess emission at lower energies. 
The obscuration scenario explains reasonably well the multi-epoch broad-band
X-ray spectra 
of the three quasars.
}
\end{figure*}

Given the limited X-ray data quality and the simplifications/assumptions made 
during the modeling process, we do not consider these spectral fitting results
fully accurate descriptions of the absorber properties. Nevertheless, they provide 
important clues for explaining
the unusual X-ray properties under our proposed obscuration-only scenario.
The best-fit results are consistent with our qualitative expectation above.
The multi-epoch spectra of PG~1001 are explained by
heavy or even Compton-thick obscuration.
A strong leaked component is required to explain its 2013 \nustar\ spectrum
as the spectral shape is likely steeper than an absorbed power law 
($\Gamma_{\rm eff}>1.5$), while
the 2020 \nustar\ spectrum is affected by typical Compton-thick obscuration.
The \chandra\ spectrum of PG~1254 requires very Compton-thick obscuration
due to the significant weakness of this $z\approx1$ spectrum.
The \nustar\ observations of PG~1254 are explained by heavy but
Compton-thin obscuration.
PHL~1811 was almost always affected by Compton-thick obscuration, and
the emergent spectra are largely dominated 
by the leaked component (though with small
$f_{\rm leak}$ values). The reprocessed component does not contribute much as
the absorber appears to have
a large covering factor ($\cos\theta_{\mbox{\scriptsize oa}}\approx0.6$)
which blocks direct reflected radiation from the opposite side of the absorber
(e.g., see the top dotted curve in Figure 6). Therefore, PHL~1811 can reach a very large
hard X-ray weakness factor ($f_{\rm w}$) in the \nustar\ observation.
For the 2015 simultaneous \xmm\ and \nustar\ observations of PHL~1811, the derived parameters
differ slightly. We also tried to tie all the parameters in
these two observations, and the results are listed in the last row of Table~3.
We do not consider the small discrepancy
a serious issue as there may be cross-calibration uncertainties between the
\xmm\ and \nustar\ data \citep[e.g.,][]{Madsen2021}.

\subsubsection{Soft X-ray Excess of PG~1001 in the Obscuration-Only Scenario}

{Soft X-ray excess emission (typically below $\approx1$~keV) is observed 
in a large fraction of type 1 AGNs, the origin
of which is still under debate and may be attributed to ionized absorption
\citep[e.g.,][]{George1998,Gierlinski2004},
ionized disk reflection \citep[e.g.,][]{Ross2005,Crummy2006}, 
or Comptonization in a warm corona \citep[e.g.,][]{Done2012}. 
Of the three quasars in this study, only PG~1001 shows a clear soft-excess
component in both its \xmm\ and \chandra\ spectra. We thus did not consider
the soft excess in the above modeling, excluding the the $<1$~keV data
for PG~1001. We explore here if the soft X-ray excess emission of PG~1001
can be explained in the obscuration-only scenario. We focus on the \xmm\
spectrum in the following discussion, 
as the \chandra\ spectrum has only 19 
counts in the 0.3--8 keV band. Nevertheless, 
we verified that the \chandra\ spectrum
yields consistent results.

The soft-excess emission of PG~1001 is at a comparable flux level to 
the $>1$~keV power-law 
component that is significantly weak compared to the expectation
from the 
$\alpha_{\rm OX}$--$L_{\rm 2500~{\textup{\AA}}}$ relation (Section 3.1.1). 
Therefore, the soft excess is also significantly weak compared to typical
levels. 
In the obscuration-only scenario, PG~1001 has a nominal-strength hot corona, 
and 
likely also a nominal-strength warm corona. 
A natural interpretation would then be that the soft-excess emission
is also filtered by the absorber if it is from the warm corona. 
We thus fitted the 0.3--10 keV \xmm\ spectrum with the same model 
described in Section 4.2 above plus an additional 
component to describe the soft excess from
the warm corona. We tested simple power-law (\texttt{zpow}), disk 
multi-black body (\texttt{diskbb}), or Comptonization (\texttt{compTT}) models
for this soft-excess component, and the three choices were all able to 
describe the spectrum well with comparable statistics. 
In Figure~8, we show the best-fit
results with the power-law model. The soft excess has a large photon 
index ($\approx5.0$) and a small normalization ($\approx0.7\%$ of the 
normalization for the intrinsic $>1$~keV power-law continuum), and the other
free parameters ($f_{\rm leak}$ and $N_{\rm H}$) are consistent with those
in Table 3 (first row) within the errors. One interpretation is thus
that the observed soft excess is the leaked portion of the 
warm corona emission through the same dust-free absorber, and the 
leaked fraction is similar to or even the same as that
for the main component. 

The soft X-ray excess emission of PG~1001 has also been suggested to be
due to ionized absorption \citep{Schartel2005}. We
verified that the 0.3--10 keV \xmm\ spetrum can be acceptably fitted with a 
simple partial-covering ionized absorption model (\texttt{zxipcf*zpow}),
fixing $\Gamma=2.6$ and the power-law normalization at 
the X-ray nominal value from
the $\alpha_{\rm OX}$--$L_{\rm 2500~{\textup{\AA}}}$ relation.
The resulting ionization parameter is $\xi\approx91$~erg~cm~s$^{-1}$ with 
$N_{\rm H}\approx4.7\times10^{23}$~cm$^{-2}$ and a covering
fraction of $\approx99.1\%$. Replacing the neutral absorption (\texttt{zphabs})
in the Section 4.2 model with \texttt{zxipcf} yields consistent results,
as the reprocessed component (\texttt{borus02}) is not important in the 
\xmm\ spectrum. The soft excess can thus also be explained with 
ionized absorption, which is possible considering that the absorber 
(e.g., disk wind) is probably partially ionized. Overall, we consider that
our proposed obscuration-only scenario can plausibly explain
the soft X-ray excess emission of PG~1001.

\begin{figure}
\includegraphics[trim=0 10 20 20,clip, width=1\linewidth]{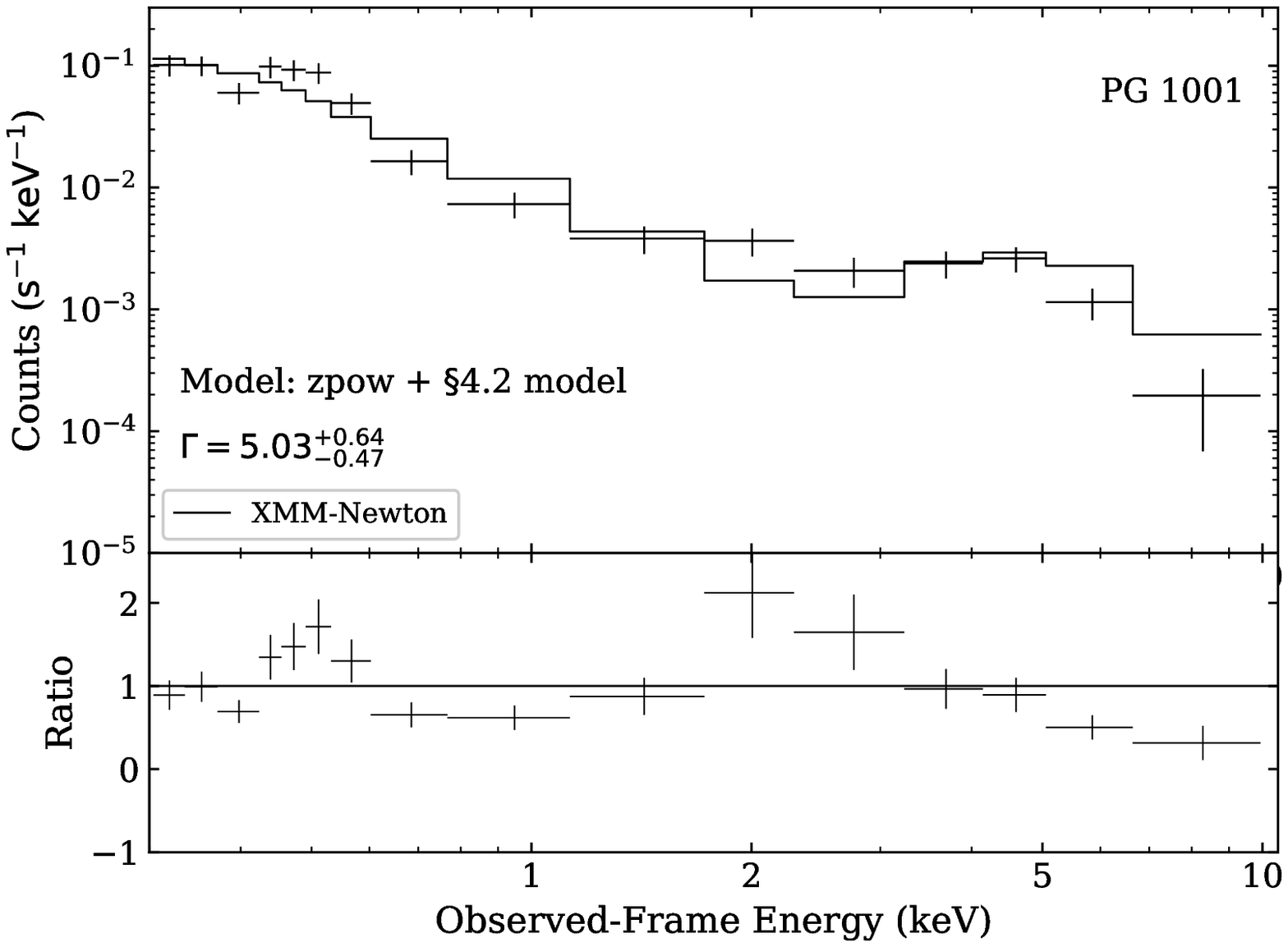}
\caption{The 0.3--10~keV \xmm\ spectrum of PG~1001 fitted with
the Section 4.2 model plus an additional power-law component.
The spectrum is grouped for display purposes only. The bottom panel displays 
the ratios of the data to the best-fit model. 
The additional power-law component describes well the
soft-excess emission with a large photon
index of $\approx5.0$. The soft excess is probably the leaked portion of the
warm corona emission through the same dust-free absorber that obscures the
$>1$~keV main component.
}
\end{figure}

}
\subsection{Clumpy Accretion-Disk Wind as the Absorber and Implications}

Under our proposed obscuration scenario above,
PG~1001 and PG~1254 are BAL quasars with intrinsically normal \xray\ emission. 
Thus, they are probably similar to the typical BAL quasars that generally show X-ray obscuration.
Besides the few low-redshift BAL quasars with \nustar\ observations \citep{Luo2013,Luo2014,Teng2014}, 
a few high-redshift BAL quasars have been suggested to be intrinsically X-ray weak
\citep{Liu2018}. 
From the systematic \chandra\ survey of the
29 high-ionization LBQS BAL quasars at
$z\approx2$, two {intrinsically \xray\ weak} candidates were
identified based on their nominal spectral shapes ($\Gamma_{\rm eff}\approx2$)
and significant hard X-ray weakness factors ($f_{\rm w}\approx12$--15);
at these redshifts, \chandra\ observations were able to provide
rest-frame hard \xray\ constraints over a similar band
to the \nustar\ observations of low-redshift objects.
We recently obtained a long
\xmm\ observation for one of the candidates, LBQS $1442-0011$. The observation was heavily affected
by background flares and the cleaned exposure time is only 40\% of the total. The results 
are summarized in the Appendix. The \xmm\ observation
suggests that the spectral shape 
became flatter ($\Gamma_{\rm eff}\approx1$) and the factor of 
hard X-ray weakness ($f_{\rm w}$) also dropped
to $\approx4$. We thus consider that the \xray\ weakness of 
this quasar can also be described by our proposed scenario of
variable obscuration, without invoking intrinsic \xray\ weakness. 
It is natural to consider that the \xray\ absorbers in PG~1001, PG~1254, and 
LBQS~$1442-0011$ are similar to those in
typical BAL quasars; i.e., the shielding gas or clumpy accretion-disk wind. 
They might
have an extreme version of the absorber in terms of its high 
column density and/or large covering factor.

An obscuration explanation for the X-ray weakness of PHL~1811 would connect it
with the PHL~1811 analogs studied in \citet{Luo2015}. PHL~1811 and its analogs belong
to a broader category of quasars, 
weak emission-line quasars (WLQs),  
a small population of type~1 quasars that show unusually weak UV emission lines
(e.g., \iona{C}{iv}). 
Systematic X-ray surveys of WLQ samples have revealed that
a large fraction ($\approx30\%$--50\%) of them are X-ray weak
\citep[e.g.,][]{Luo2015,Ni2018,Ni2022,Pu2020}.
These WLQs
have typically been selected to lie at $z\gtrsim1.5$, and thus 
\chandra\ or \xmm\ observations provide rest-frame hard X-ray constraints.
The individual and stacked effective power-law photon indices for the X-ray weak WLQs 
are in general flat ($\Gamma_{\rm eff}\approx1.2$),
suggesting an obscuration scenario.  
The absorber is proposed to be the geometrically thick inner accretion disk and/or its 
associated disk wind, which also shields the nuclear EUV/X-ray ionizing radiation from
reaching the BELR, causing the weak emission lines.
Thick inner accretion disks are expected in WLQs as they are considered to have high
or even super-Eddington accretion rates that result in thick disks 
\citep[e.g.,][]{Abramowicz1988,Mineshige2000,Wang2003,Ohsuga2011,Jiang2014}.
Powerful accretion-disk winds launched via radiation pressure are
expected in such systems \citep[e.g.,][]{Takeuchi2014,Jiang2019,Giustini2019}.
PHL~1811 likely has a super-Eddington accretion rate, and the large covering-factor absorber
is probably the strong wind associated with a thick accretion disk.
 
Therefore, the three quasars in this study likely share a similar nature, with 
partial-covering absorption
from clumpy dust-free winds (e.g., Figure 6). 
X-ray absorption from clumpy winds/outflows has been observed in typical
type~1 AGNs \citep[e.g.,][]{Kaastra2014,Mehdipour2017,Mehdipour2021,Dehghanian2019,Laha2021},
although the absorption strength is often not comparable to the extreme X-ray weakness found in 
our three quasars. The wind strength and density likely have an 
Eddington ratio dependence, 
as heavy or even Compton-thick absorption has been observed in
super-Eddington accreting AGNs \citep[e.g.,][]{Longinotti2019,Liu2021}.
The three quasars in this study 
probably have super-Eddington accretion rates that drive powerful and high-density 
clumpy winds.
They have large 
estimated Eddington ratios as listed in Table~1.
We note that these Eddington ratios
do not represent accurately the
accretion power in the super-Eddington regime as a large
fraction of the power may be 
advected into the SMBH or converted into mechanical
energy of the wind \citep[e.g.,][]{Jiang2019}; also see Section~2.1 for further discussion.
The large intrinsic X-ray photon indices derived from spectral fitting 
($\Gamma\approx2.1$--2.6; Table~3)
also suggest super-Eddington accretion rates \citep[e.g.,][]{Shemmer2008,Huang2020}.
The NLQ1 classification and the weak [\iona{O}{iii}] emission of PG~1001 and PHL~1811 
(Section 2.1)
provide additional support of super-Eddington accretion in these two quasars \citep[e.g.,][]{Boroson1992,Sulentic2000,Shen2014}.

It is somewhat odd that PG~1001 shows a strong \iona{C}{iv} emission line in its UV spectrum
\citep{Brandt2000},
as the strong and high-density wind and/or the
thick inner accretion disk
should be able to shield
the BELR from the nuclear ionization, resulting in a WLQ like PHL~1811. 
It also appears unusual that PHL~1811 does not show any significant UV absorption lines (i.e., BALs).
Perhaps the dynamical nature of the wind (e.g., variable $N_{\rm H}$ and covering factor) 
causes the apparent discrepancy, and multi-epoch UV spectra might be able to shed some light.
For example, a $z\approx2$ WLQ has recently been found to undergo BAL transformation 
\citep{Yi2022}.
Geometric effects might also play a role, as the line of sight to the X-ray corona, 
line of sight to the accretion-disk UV continuum region, and the direction from the nucleus to
the BELR are different from each other, and thus the emergent X-ray and UV spectra 
depend on the physical configuration of the clumpy wind \citep[e.g.,][]{Giustini2019}. 

Although our proposed obscuration scenario was based on the
new sensitive \nustar\ observations of these 
three quasars, the general connection with obscuration from the disk wind suggests 
that this scenario may be applicable to the other intrinsically 
\xray\ weak quasar candidates (e.g., those in \citealt{Nardini2019} and
\citealt{Laurenti2021}). 
Obscuration from the clumpy disk wind would predict 
X-ray variability from varying obscuration. 
For the three quasars
in this study, PG~1001 and PHL~1811 showed clear soft X-ray variability (e.g., Figure~1); 
the \hbox{12-day} variability timescale of PHL~1811
does not provide any strong constraints on the wind velocity as a wind clump only needs to
move a fraction of the corona size. 
The \nustar\ observations 
of PG~1001 suggest some hard \xray\ variability, at least in the spectral shape (Table~2).
The PG~1254 \nustar\ observations do not provide sufficient photon 
statistics to identify hard \xray\ variability.
In addition, LBQS~$1442-0011$ likely has hard X-ray variability
(see the Appendix).
The WLQs have limited multi-epoch observations, and
a few of them have been found to vary strongly between X-ray normal and \xray\ weak states
\citep{Miniutti2012,Ni2020}. A small fraction of super-Eddington accreting AGNs have also
been found to vary between X-ray normal and X-ray weak states 
\citep[e.g.,][]{Liu2019,Boller2021,Liu2021}, and a few also show steep spectra in the low state.
{Another characteristic property of such \hbox{X-ray} variability is that there
is no contemporaneous optical/UV continuum or emission-line variability, which argues against
changes of accretion rates and supports the obscuration scenario.
This property also makes these AGNs distinct from the unusual population of
``changing-look'' AGNs (e.g., 1ES~$1927+654$; \citealt{Trakhtenbrot2019,Ricci2021}) 
that also show extreme X-ray variability but are generally 
attributed to 
changes of accretion rates or tidal-disruption events.}
Multi-epoch X-ray observations of the 
intrinsically \xray\ weak quasar candidates might be able to
reveal X-ray variability and help clarify
their nature. 

\section{Summary and Future Work}
In this paper, we used \nustar\ observations of 
PG~1001, PG~1254, and PHL~1811 to constrain their hard \xray\ (\hbox{$\gtrsim5$~keV}) weakness and spectral
shapes, and thus
to investigate the nature of their extreme X-ray weakness.
These quasars show very weak soft X-ray emission (Figure 1), and they
were previously proposed to be intrinsically X-ray weak, with the X-ray coronae
producing weak continuum emission relative to their optical/UV emission (deviating
below the $\alpha_{\rm OX}$--$L_{\rm 2500~{\textup{\AA}}}$ relation).
The multi-epoch soft and hard X-ray observations are summarized in Table 1.
\nustar\ aperture photometry was presented in Section 3.2, and the results 
are summarized in Table 2 and Figure 3. The \nustar\ spectral shapes for PG~1001 and 
PHL~1811 appear flat ($\Gamma_{\rm eff}=1.0^{+0.5}_{-0.6}$ and 
$\Gamma_{\rm eff}=1.4^{+0.8}_{-0.7}$, respectively), while the shape is 
nominal
for PG~1254 ($\Gamma_{\rm eff}=1.8\pm0.3$). 
PG~1001 and PHL~1811 are significantly hard X-ray weak
compared to the expectations from their optical/UV emission ($f_{\rm w}$ at 8~keV 
$\approx26$--74), while PG~1254 is only X-ray weak
by a factor of $\approx3$. 
The PHL~1811 hard X-ray 
photon index appears smaller than
its soft X-ray (0.3--5 keV) photon index ($2.3\pm0.1$).
Spectral modeling suggests that its 2015 \xmm\ and \nustar\ spectra cannot be 
described by an intrinsically weak continuum plus a reasonable amount of 
Compton reflection (Section 4.1 and Figures 4 \& 5).

In light of the new \nustar\ results, a variable X-ray absorber can account for 
all the observations of these X-ray weak quasars.
We propose that, as an alternative to the 
intrinsic X-ray weakness $+$ X-ray obscuration scenario, 
the soft and hard X-ray weakness of these quasars can be uniformly explained 
under an X-ray obscuration-only
scenario, without invoking the 
extra mechanism of intrinsic X-ray weakness (Section 4.2). 
In this scenario, the weak emergent spectrum is 
a combination of the transmitted component modified by absorption,
the leaked component through a clumpy absorber (including a distant scattered 
component), and the
reprocessed component reflected/scattered from the absorber (Figure~6). 
This 
partial-covering absorption scenario 
provides adequate explanations of the 
multi-epoch X-ray data of these quasars, and the X-ray variability is mainly
induced by the varying column density and leaked fraction (partial-covering fraction) 
of the absorber (Table 3 and Figure 7).

We propose that the absorber is a clumpy dust-free wind launched from the accretion disk 
(Section 4.3).
These quasars probably have super-Eddington accretion 
rates which result in 
geometrically thick inner accretion disks and powerful winds with 
high column densities and large covering factors.
Although we cannot rigorously prove that intrinsic \xray\ weakness is not present in
these systems,
the connections of these quasars to other X-ray weak quasars including WLQs and 
super-Eddington accreting quasars
point to a universal wind obscuration scenario 
for the weak X-ray emission found in type 1 quasars, or even type 1 AGNs in general.
Multi-epoch X-ray observations of the
intrinsically X-ray weak quasar candidates will further help clarify
their nature. Besides variability investigations, deeper \nustar\ 
observations of PHL~1811 could provide further evidence of heavy X-ray obscuration.
Also, higher signal-to-noise ratio and higher spectral resolution observations 
with future generation X-ray
observatories (e.g., Athena; \citealt{Nandra2013}) 
could reveal spectral features (e.g., the Fe lines in the reprocessed component) 
that help discriminate between 
different scenarios.

~\\

C.W. and B.L. acknowledge financial support from
the National Natural Science Foundation of China
grant 11991053, 
China Manned Space Project grants NO. CMS-CSST-2021-A05
and NO. CMS-CSST-2021-A06.
W.N.B. acknowledges support from the V.M. Willaman Endowment, NASA grants 80NSSC20K0029 and 80NSSC22K0071, and Penn State ACIS Instrument Team Contract SV4-74018 (issued by the Chandra X-ray Center, which is operated by the Smithsonian Astrophysical Observatory for and on behalf of NASA under contract NAS8-03060).
F.E.B acknowledges support from ANID-Chile BASAL AFB-170002 and FB210003, FONDECYT Regular 1200495 and 1190818, and Millennium Science Initiative Program  – ICN12\_009.
S.C.G thanks the Natural Science and Engineering Research Council of Canada.

We have made use of data from the NuSTAR mission, a project led by the California Institute of Technology, managed by the Jet Propulsion Laboratory and funded by the National Aeronautics and Space Administration. We thank the NuSTAR Operations, Software and Calibration teams for support with the execution and analysis of these observations. This research has made use of the NuSTAR Data Analysis Software (NuSTARDAS) jointly developed by the ASI Science Data Center (ASDC, Italy) and the California Institute of Technology (USA).

The Chandra ACIS Team Guaranteed Time Observations (GTO) utilized were selected by the ACIS Instrument Principal Investigator, Gordon P. Garmire, currently of the Huntingdon Institute for X-ray Astronomy, LLC, which is under contract to the Smithsonian Astrophysical Observatory via Contract SV2-82024.


\begin{thebibliography}{}
\expandafter\ifx\csname natexlab\endcsname\relax\def\natexlab#1{#1}\fi
\providecommand{\url}[1]{\href{#1}{#1}}
\providecommand{\dodoi}[1]{doi:~\href{http://doi.org/#1}{\nolinkurl{#1}}}
\providecommand{\doeprint}[1]{\href{http://ascl.net/#1}{\nolinkurl{http://ascl.net/#1}}}
\providecommand{\doarXiv}[1]{\href{https://arxiv.org/abs/#1}{\nolinkurl{https://arxiv.org/abs/#1}}}

\bibitem[{{Abramowicz} {et~al.}(1988){Abramowicz}, {Czerny}, {Lasota}, \&
  {Szuszkiewicz}}]{Abramowicz1988}
{Abramowicz}, M.~A., {Czerny}, B., {Lasota}, J.~P., \& {Szuszkiewicz}, E. 1988,
  \apj, 332, 646

\bibitem[{{Arnaud}(1996)}]{Arnaud1996}
{Arnaud}, K.~A. 1996, in \aspc, Vol. 101, Astronomical Data Analysis Software
  and Systems V, ed. G.~H. {Jacoby} \& J.~{Barnes}, 17

\bibitem[{{Balokovi{\'c}} {et~al.}(2018){Balokovi{\'c}}, {Brightman},
  {Harrison}, {Comastri}, {Ricci}, {Buchner}, {Gandhi}, {Farrah}, \&
  {Stern}}]{Balokovi2018}
{Balokovi{\'c}}, M., {Brightman}, M., {Harrison}, F.~A., {et~al.} 2018, \apj,
  854, 42

\bibitem[{{Baskin} {et~al.}(2014){Baskin}, {Laor}, \& {Stern}}]{Baskin2014}
{Baskin}, A., {Laor}, A., \& {Stern}, J. 2014, \mnras, 445, 3025

\bibitem[{{Bellm} {et~al.}(2019){Bellm}, {Kulkarni}, {et~al.}}]{Bellm2019}
{Bellm}, E.~C., {Kulkarni}, S.~R., {et~al.} 2019, \pasp, 131, 018002

\bibitem[{{Boller} {et~al.}(2021){Boller}, {Liu}, {Weber},
  {et~al.}}]{Boller2021}
{Boller}, T., {Liu}, T., {Weber}, P., {et~al.} 2021, \aap, 647, A6

\bibitem[{{Boroson} \& {Green}(1992)}]{Boroson1992}
{Boroson}, T.~A., \& {Green}, R.~F. 1992, \apjs, 80, 109

\bibitem[{{Brandt} {et~al.}(2000){Brandt}, {Laor}, \& {Wills}}]{Brandt2000}
{Brandt}, W.~N., {Laor}, A., \& {Wills}, B.~J. 2000, \apj, 528, 637

\bibitem[{{Cappi} {et~al.}(2006){Cappi}, {Panessa}, {Bassani}, {Dadina}, {Di
  Cocco}, {Comastri}, {della Ceca}, {Filippenko}, {Gianotti}, {Ho}, {Malaguti},
  {Mulchaey}, {Palumbo}, {Piconcelli}, {Sargent}, {Stephen}, {Trifoglio}, \&
  {Weaver}}]{Cappi2006}
{Cappi}, M., {Panessa}, F., {Bassani}, L., {et~al.} 2006, \aap, 446, 459

\bibitem[{{Cardelli} {et~al.}(1989){Cardelli}, {Clayton}, \&
  {Mathis}}]{Cardelli1989}
{Cardelli}, J.~A., {Clayton}, G.~C., \& {Mathis}, J.~S. 1989, \apj, 345, 245

\bibitem[{{Chen} {et~al.}(2017){Chen}, {Hickox}, {Goulding}, {Stern}, {Assef},
  {Kochanek}, {Brown}, {Harrison}, {Hainline}, {Alberts}, {Alexander},
  {Brodwin}, {Del Moro}, {Forman}, {Gorjian}, {Jones}, {Murray}, {Pope}, \&
  {Rovilos}}]{Chen2017}
{Chen}, C.-T.~J., {Hickox}, R.~C., {Goulding}, A.~D., {et~al.} 2017, \apj, 837,
  145

\bibitem[{{Comastri} {et~al.}(2011){Comastri}, {Ranalli}, {Iwasawa},
  {et~al.}}]{Comastri2011}
{Comastri}, A., {Ranalli}, P., {Iwasawa}, K., {et~al.} 2011, \aap, 526, L9

\bibitem[{{Crummy} {et~al.}(2006){Crummy}, {Fabian}, {Gallo}, \&
  {Ross}}]{Crummy2006}
{Crummy}, J., {Fabian}, A.~C., {Gallo}, L., \& {Ross}, R.~R. 2006, \mnras, 365,
  1067

\bibitem[{{Dauser} {et~al.}(2014){Dauser}, {Garcia}, {Parker}, {Fabian}, \&
  {Wilms}}]{Dauser2014}
{Dauser}, T., {Garcia}, J., {Parker}, M.~L., {Fabian}, A.~C., \& {Wilms}, J.
  2014, \mnras, 444, L100

\bibitem[{{de Rosa} {et~al.}(2008){de Rosa}, {Bassani}, {Ubertini}, {Malizia},
  {Dean}, \& {Walter}}]{deRosa2008}
{de Rosa}, A., {Bassani}, L., {Ubertini}, P.~{Panessa}, F., {et~al.} 2008,
  \aap, 483, 749

\bibitem[{{Dehghanian} {et~al.}(2019){Dehghanian}, {Ferland}, {Kriss},
  {Peterson}, {Mathur}, {Mehdipour}, {Guzm{\'a}n}, {Chatzikos}, {van Hoof},
  {Williams}, {Arav}, {Barth}, {Bentz}, {Bisogni}, {Brandt}, {Crenshaw}, {Dalla
  Bont{\`a}}, {De Rosa}, {Fausnaugh}, {Gelbord}, {Goad}, {Gupta}, {Horne},
  {Kaastra}, {Knigge}, {Korista}, {McHardy}, {Pogge}, {Starkey}, \&
  {Vestergaard}}]{Dehghanian2019}
{Dehghanian}, M., {Ferland}, G.~J., {Kriss}, G.~A., {et~al.} 2019, \apj, 877,
  119

\bibitem[{{Done}(2010)}]{Done2010}
{Done}, C. 2010, arXiv e-prints, arXiv:1008.2287.
\newblock \doarXiv{1008.2287}

\bibitem[{{Done} {et~al.}(2012){Done}, {Davis}, {Jin}, {Blaes}, \&
  {Ward}}]{Done2012}
{Done}, C., {Davis}, S.~W., {Jin}, C., {Blaes}, O., \& {Ward}, M. 2012, \mnras,
  420, 1848

\bibitem[{{Drake} {et~al.}(2009){Drake}, {Djorgovski}, {Mahabal},
  {et~al.}}]{Drake2009}
{Drake}, A.~J., {Djorgovski}, S.~G., {Mahabal}, A., {et~al.} 2009, \apj, 696,
  870

\bibitem[{{Fabian} {et~al.}(2017){Fabian}, {Alston}, {Cackett}, {Kara},
  {Uttley}, \& {Wilkins}}]{Fabian2017}
{Fabian}, A.~C., {Alston}, W.~N., {Cackett}, E.~M., {et~al.} 2017,
  Astronomische Nachrichten, 338, 269

\bibitem[{{Fabian} {et~al.}(2013){Fabian}, {Kara}, {Walton}, {Wilkins}, {Ross},
  {Lozanov}, {Uttley}, {Gallo}, {Zoghbi}, {Miniutti}, {Boller}, {Brandt},
  {Cackett}, {Chiang}, {Dwelly}, {Malzac}, {Miller}, {Nardini}, {Ponti},
  {Reis}, {Reynolds}, {Steiner}, {Tanaka}, \& {Young}}]{Fabian2013}
{Fabian}, A.~C., {Kara}, E., {Walton}, D.~J., {et~al.} 2013, \mnras, 429, 2917

\bibitem[{{Fan} {et~al.}(2009){Fan}, {Wang}, {Wang}, {Wang}, {Dong}, {Zhang},
  \& {Cheng}}]{Fan2009}
{Fan}, L.~L., {Wang}, H.~Y., {Wang}, T., {et~al.} 2009, \apj, 690, 1006

\bibitem[{{Freeman} {et~al.}(2002){Freeman}, {Kashyap}, {Rosner}, \&
  {Lamb}}]{Freeman2002}
{Freeman}, P.~E., {Kashyap}, V., {Rosner}, R., \& {Lamb}, D.~Q. 2002, \apjs,
  138, 185

\bibitem[{{Gallagher} {et~al.}(2002){Gallagher}, {Brandt}, {Chartas}, \&
  {Garmire}}]{Gallagher2002}
{Gallagher}, S.~C., {Brandt}, W.~N., {Chartas}, G., \& {Garmire}, G.~P. 2002,
  \apj, 567, 37

\bibitem[{{Gallagher} {et~al.}(2006){Gallagher}, {Brandt}, {Chartas},
  {Priddey}, {Garmire}, \& {Sambruna}}]{Gallagher2006}
{Gallagher}, S.~C., {Brandt}, W.~N., {Chartas}, G., {et~al.} 2006, \apj, 644,
  709

\bibitem[{{Gandhi} {et~al.}(2014){Gandhi}, {Lansbury}, {Alexander}, {Stern},
  {Ar{\'e}valo}, {Ballantyne}, {Balokovi{\'c}}, {Bauer}, {Boggs}, {Brandt},
  {Brightman}, {Christensen}, {Comastri}, {Craig}, {Del Moro}, {Elvis},
  {Fabian}, {Hailey}, {Harrison}, {Hickox}, {Koss}, {LaMassa}, {Luo},
  {Madejski}, {Ptak}, {Puccetti}, {Teng}, {Urry}, {Walton}, \&
  {Zhang}}]{Gandhi2014}
{Gandhi}, P., {Lansbury}, G.~B., {Alexander}, D.~M., {et~al.} 2014, \apj, 792,
  117

\bibitem[{{Garc{\'\i}a} {et~al.}(2014){Garc{\'\i}a}, {Dauser}, {Lohfink},
  {Kallman}, {Steiner}, {McClintock}, {Brenneman}, {Wilms}, {Eikmann},
  {Reynolds}, \& {Tombesi}}]{Garcia2014}
{Garc{\'\i}a}, J., {Dauser}, T., {Lohfink}, A., {et~al.} 2014, \apj, 782, 76

\bibitem[{{Gehrels}(1986)}]{Gehrels1986}
{Gehrels}, N. 1986, \apj, 303, 336

\bibitem[{{George} \& {Fabian}(1991)}]{George1991}
{George}, I.~M., \& {Fabian}, A.~C. 1991, \mnras, 249, 352

\bibitem[{{George} {et~al.}(1998){George}, {Turner}, {Netzer}, {Nandra},
  {Mushotzky}, \& {Yaqoob}}]{George1998}
{George}, I.~M., {Turner}, T.~J., {Netzer}, H., {et~al.} 1998, \apjs, 114, 73

\bibitem[{{Gibson} {et~al.}(2009){Gibson}, {Jiang}, {Brandt}, {Hall}, {Shen},
  {Wu}, {Anderson}, {Schneider}, {Vanden Berk}, {Gallagher}, {Fan}, \&
  {York}}]{Gibson2009}
{Gibson}, R.~R., {Jiang}, L., {Brandt}, W.~N., {et~al.} 2009, \apj, 692, 758

\bibitem[{{Gierli{\'n}ski} \& {Done}(2004)}]{Gierlinski2004}
{Gierli{\'n}ski}, M., \& {Done}, C. 2004, \mnras, 349, L7

\bibitem[{{Gilfanov} \& {Merloni}(2014)}]{Gilfanov2014}
{Gilfanov}, M., \& {Merloni}, A. 2014, \ssr, 183, 121

\bibitem[{{Giustini} \& {Proga}(2019)}]{Giustini2019}
{Giustini}, M., \& {Proga}, D. 2019, \aap, 630, A94

\bibitem[{{Gupta} {et~al.}(2021){Gupta}, {Ricci}, {Tortosa}, {Ueda},
  {Kawamuro}, {Koss}, {Trakhtenbrot}, {Oh}, {Bauer}, {Ricci}, {Privon},
  {Zappacosta}, {Stern}, {Kakkad}, {Piconcelli}, {Veilleux}, {Mushotzky},
  {Caglar}, {Ichikawa}, {Elagali}, {Powell}, {Urry}, \& {Harrison}}]{Gupta2021}
{Gupta}, K.~K., {Ricci}, C., {Tortosa}, A., {et~al.} 2021, \mnras, 504, 428

\bibitem[{{HI4PI Collaboration} {et~al.}(2016){HI4PI Collaboration}, {Ben
  Bekhti}, {Fl{\"o}er}, {et~al.}}]{HI4PI2016}
{HI4PI Collaboration}, {Ben Bekhti}, N., {Fl{\"o}er}, L., {et~al.} 2016, \aap,
  594, A116

\bibitem[{{Hickox} \& {Alexander}(2018)}]{Hickox2018}
{Hickox}, R.~C., \& {Alexander}, D.~M. 2018, \araa, 56, 625

\bibitem[{{Huang} {et~al.}(2020){Huang}, {Luo}, {Du}, {Hu}, {Wang}, \&
  {Li}}]{Huang2020}
{Huang}, J., {Luo}, B., {Du}, P., {et~al.} 2020, \apj, 895, 114

\bibitem[{{Immler} {et~al.}(2003){Immler}, {Brandt}, {Vignali}, {Bauer},
  {Crenshaw}, {Feldmeier}, \& {Kraemer}}]{Immler2003}
{Immler}, S., {Brandt}, W.~N., {Vignali}, C., {et~al.} 2003, \aj, 126, 153

\bibitem[{{Jiang} {et~al.}(2014){Jiang}, {Stone}, \& {Davis}}]{Jiang2014}
{Jiang}, Y.-F., {Stone}, J.~M., \& {Davis}, S.~W. 2014, \apj, 796, 106

\bibitem[{{Jiang} {et~al.}(2019){Jiang}, {Stone}, \& {Davis}}]{Jiang2019}
---. 2019, \apj, 880, 67

\bibitem[{{Just} {et~al.}(2007){Just}, {Brandt}, {Shemmer}, {Steffen},
  {Schneider}, {Chartas}, \& {Garmire}}]{Just2007}
{Just}, D.~W., {Brandt}, W.~N., {Shemmer}, O., {et~al.} 2007, \apj, 665, 1004

\bibitem[{{Kaastra} {et~al.}(2014){Kaastra}, {Kriss}, {Cappi}, {Mehdipour},
  {Petrucci}, {Steenbrugge}, {Arav}, {Behar}, {Bianchi}, {Boissay},
  {Branduardi-Raymont}, {Chamberlain}, {Costantini}, {Ely}, {Ebrero}, {Di
  Gesu}, {Harrison}, {Kaspi}, {Malzac}, {De Marco}, {Matt}, {Nandra},
  {Paltani}, {Person}, {Peterson}, {Pinto}, {Ponti}, {Nu{\~n}ez}, {De Rosa},
  {Seta}, {Ursini}, {de Vries}, {Walton}, \& {Whewell}}]{Kaastra2014}
{Kaastra}, J.~S., {Kriss}, G.~A., {Cappi}, M., {et~al.} 2014, Science, 345, 64

\bibitem[{{Kraft} {et~al.}(1991){Kraft}, {Burrows}, \& {Nousek}}]{Kraft1991}
{Kraft}, R.~P., {Burrows}, D.~N., \& {Nousek}, J.~A. 1991, \apj, 374, 344

\bibitem[{{Krawczyk} {et~al.}(2013){Krawczyk}, {Richards}, {Mehta}, {Vogeley},
  {Gallagher}, {Leighly}, {Ross}, \& {Schneider}}]{Krawczyk2013}
{Krawczyk}, C.~M., {Richards}, G.~T., {Mehta}, S.~S., {et~al.} 2013, \apjs,
  206, 4

\bibitem[{{Laha} {et~al.}(2021){Laha}, {Reynolds}, {Reeves}, {Kriss},
  {Guainazzi}, {Smith}, {Veilleux}, \& {Proga}}]{Laha2021}
{Laha}, S., {Reynolds}, C.~S., {Reeves}, J., {et~al.} 2021, Nature Astronomy,
  5, 13

\bibitem[{{Lansbury} {et~al.}(2017){Lansbury}, {Stern}, {Aird}, {Alexander},
  {Fuentes}, {Harrison}, {Treister}, {Bauer}, {Tomsick}, {Balokovi{\'c}}, {Del
  Moro}, {Gandhi}, {Ajello}, {Annuar}, {Ballantyne}, {Boggs}, {Brandt},
  {Brightman}, {Chen}, {Christensen}, {Civano}, {Comastri}, {Craig}, {Forster},
  {Grefenstette}, {Hailey}, {Hickox}, {Jiang}, {Jun}, {Koss}, {Marchesi},
  {Melo}, {Mullaney}, {Noirot}, {Schulze}, {Walton}, {Zappacosta}, \&
  {Zhang}}]{Lansbury2017}
{Lansbury}, G.~B., {Stern}, D., {Aird}, J., {et~al.} 2017, \apj, 836, 99

\bibitem[{{Laurenti} {et~al.}(2022){Laurenti}, {Piconcelli}, {Zappacosta},
  {Tombesi}, {Vignali}, {Bianchi}, {Marziani}, {Vagnetti}, {Bongiorno},
  {Bischetti}, {del Olmo}, {Lanzuisi}, {Luminari}, {Middei}, {Perri}, {Ricci},
  \& {Vietri}}]{Laurenti2021}
{Laurenti}, M., {Piconcelli}, E., {Zappacosta}, L., {et~al.} 2022, \aap, 657,
  A57

\bibitem[{{Leighly} {et~al.}(2007{\natexlab{a}}){Leighly}, {Halpern},
  {Jenkins}, \& {Casebeer}}]{Leighly2007-2}
{Leighly}, K.~M., {Halpern}, J.~P., {Jenkins}, E.~B., \& {Casebeer}, D.
  2007{\natexlab{a}}, \apjs, 173, 1

\bibitem[{{Leighly} {et~al.}(2007{\natexlab{b}}){Leighly}, {Halpern},
  {Jenkins}, {Grupe}, {Choi}, \& {Prescott}}]{Leighly2007}
{Leighly}, K.~M., {Halpern}, J.~P., {Jenkins}, E.~B., {et~al.}
  2007{\natexlab{b}}, \apj, 663, 103

\bibitem[{{Leighly} {et~al.}(2019){Leighly}, {Terndrup}, {Lucy}, {Choi},
  {Gallagher}, {Richards}, {Dietrich}, \& {Raney}}]{Leighly2019}
{Leighly}, K.~M., {Terndrup}, D.~M., {Lucy}, A.~B., {et~al.} 2019, \apj, 879,
  27

\bibitem[{{Liu} {et~al.}(2021){Liu}, {Luo}, {Brandt}, {Brotherton},
  {Gallagher}, {Ni}, {Shemmer}, \& {Timlin}}]{Liu2021}
{Liu}, H., {Luo}, B., {Brandt}, W.~N., {et~al.} 2021, \apj, 910, 103

\bibitem[{{Liu} {et~al.}(2018){Liu}, {Luo}, {Brandt}, {Gallagher}, \&
  {Garmire}}]{Liu2018}
{Liu}, H., {Luo}, B., {Brandt}, W.~N., {Gallagher}, S.~C., \& {Garmire}, G.~P.
  2018, \apj, 859, 113

\bibitem[{{Liu} {et~al.}(2019){Liu}, {Luo}, {Brandt}, {Brotherton}, {Du},
  {Gallagher}, {Hu}, {Shemmer}, \& {Wang}}]{Liu2019}
{Liu}, H., {Luo}, B., {Brandt}, W.~N., {et~al.} 2019, \apj, 878, 79

\bibitem[{{Longinotti} {et~al.}(2019){Longinotti}, {Kriss}, {Krongold},
  {Arellano-Cordova}, {Komossa}, {Gallo}, {Grupe}, {Mathur}, {Parker},
  {Pradhan}, \& {Wilkins}}]{Longinotti2019}
{Longinotti}, A.~L., {Kriss}, G., {Krongold}, Y., {et~al.} 2019, \apj, 875, 150

\bibitem[{{Luo} {et~al.}(2013){Luo}, {Brandt}, {Alexander}, {Harrison},
  {Stern}, {Bauer}, {Boggs}, {Christensen}, {Comastri}, {Craig}, {Fabian},
  {Farrah}, {Fiore}, {Fuerst}, {Grefenstette}, {Hailey}, {Hickox}, {Madsen},
  {Matt}, {Ogle}, {Risaliti}, {Saez}, {Teng}, {Walton}, \& {Zhang}}]{Luo2013}
{Luo}, B., {Brandt}, W.~N., {Alexander}, D.~M., {et~al.} 2013, \apj, 772, 153

\bibitem[{{Luo} {et~al.}(2014){Luo}, {Brandt}, {Alexander}, {Stern}, {Teng},
  {Ar{\'e}valo}, {Bauer}, {Boggs}, {Christensen}, {Comastri}, {Craig},
  {Farrah}, {Gandhi}, {Hailey}, {Harrison}, {Koss}, {Ogle}, {Puccetti}, {Saez},
  {Scott}, {Walton}, \& {Zhang}}]{Luo2014}
---. 2014, \apj, 794, 70

\bibitem[{{Luo} {et~al.}(2015){Luo}, {Brandt}, {Hall}, {Wu}, {Anderson},
  {Garmire}, {Gibson}, {Plotkin}, {Richards}, {Schneider}, {Shemmer}, \&
  {Shen}}]{Luo2015}
{Luo}, B., {Brandt}, W.~N., {Hall}, P.~B., {et~al.} 2015, \apj, 805, 122

\bibitem[{{Lusso} \& {Risaliti}(2017)}]{Lusso2017}
{Lusso}, E., \& {Risaliti}, G. 2017, \aap, 602, A79


\bibitem[{{Lutz} {et~al.}(2004){Lutz}, {Maiolino}, {Spoon}, \&
  {Moorwood}}]{Lutz2004}
{Lutz}, D., {Maiolino}, R., {Spoon}, H.~W.~W., \& {Moorwood}, A.~F.~M. 2004,
  \aap, 418, 465

\bibitem[{{Madsen} {et~al.}(2021){Madsen}, {Burwitz}, {Forster}, {Grant},
  {Guainazzi}, {Kashyap}, {Marshall}, {Miller}, {Natalucci}, {Plucinsky}, \&
  {Terada}}]{Madsen2021}
{Madsen}, K.~K., {Burwitz}, V., {Forster}, K., {et~al.} 2021, arXiv e-prints,
  arXiv:2111.01613.
\newblock \doarXiv{2111.01613}

\bibitem[{{Mainzer} {et~al.}(2014){Mainzer}, {Bauer}, {Cutri}, {Grav},
  {Masiero}, {Beck}, {Clarkson}, {Conrow}, {Dailey}, {Eisenhardt}, {Fabinsky},
  {Fajardo-Acosta}, {Fowler}, {Gelino}, {Grillmair}, {Heinrichsen}, {Kendall},
  {Kirkpatrick}, {Liu}, {Masci}, {McCallon}, {Nugent}, {Papin}, {Rice},
  {Royer}, {Ryan}, {Sevilla}, {Sonnett}, {Stevenson}, {Thompson}, {Wheelock},
  {Wiemer}, {Wittman}, {Wright}, \& {Yan}}]{Mainzer2014}
{Mainzer}, A., {Bauer}, J., {Cutri}, R.~M., {et~al.} 2014, \apj, 792, 30

\bibitem[{{Marconi} {et~al.}(2008){Marconi}, {Axon}, {Maiolino}, {Nagao},
  {Pastorini}, {Pietrini}, {Robinson}, \& {Torricelli}}]{Marconi2008}
{Marconi}, A., {Axon}, D.~J., {Maiolino}, R., {et~al.} 2008, \apj, 678, 693

\bibitem[{{Marconi} {et~al.}(2009){Marconi}, {Axon}, {Maiolino}, {Nagao},
  {Pietrini}, {Risaliti}, {Robinson}, \& {Torricelli}}]{Marconi2009}
---. 2009, \apjl, 698, L103

\bibitem[{{Martin} {et~al.}(2005){Martin}, {Fanson}, {Schiminovich},
  {Morrissey}, {Friedman}, {Barlow}, {Conrow}, {Grange}, {Jelinsky},
  {Milliard}, {Siegmund}, {Bianchi}, {Byun}, {Donas}, {Forster}, {Heckman},
  {Lee}, {Madore}, {Malina}, {Neff}, {Rich}, {Small}, {Surber}, {Szalay},
  {Welsh}, \& {Wyder}}]{Martin2005}
{Martin}, D.~C., {Fanson}, J., {Schiminovich}, D., {et~al.} 2005, \apjl, 619,
  L1

\bibitem[{{Martocchia} {et~al.}(2017){Martocchia}, {Piconcelli}, {Zappacosta},
  {Duras}, {Vietri}, {Vignali}, {Bianchi}, {Bischetti}, {Bongiorno}, {Brusa},
  {Lanzuisi}, {Marconi}, {Mathur}, {Miniutti}, {Nicastro}, {Bruni}, \&
  {Fiore}}]{Martocchia2017}
{Martocchia}, S., {Piconcelli}, E., {Zappacosta}, L., {et~al.} 2017, \aap, 608,
  A51

\bibitem[{{Mateos} {et~al.}(2015){Mateos}, {Carrera}, {Alonso-Herrero},
  {Rovilos}, {Hern{\'a}n-Caballero}, {Barcons}, {Blain}, {Caccianiga}, {Della
  Ceca}, \& {Severgnini}}]{Mateos2015}
{Mateos}, S., {Carrera}, F.~J., {Alonso-Herrero}, A., {et~al.} 2015, \mnras,
  449, 1422

\bibitem[{{Matthews} {et~al.}(2016){Matthews}, {Knigge}, {Long}, {Sim},
  {Higginbottom}, \& {Mangham}}]{Matthews2016}
{Matthews}, J.~H., {Knigge}, C., {Long}, K.~S., {et~al.} 2016, \mnras, 458, 293

\bibitem[{{Mehdipour} {et~al.}(2017){Mehdipour}, {Kaastra}, {Kriss}, {Arav},
  {Behar}, {Bianchi}, {Branduardi-Raymont}, {Cappi}, {Costantini}, {Ebrero},
  {Di Gesu}, {Kaspi}, {Mao}, {De Marco}, {Matt}, {Paltani}, {Peretz},
  {Peterson}, {Petrucci}, {Pinto}, {Ponti}, {Ursini}, {de Vries}, \&
  {Walton}}]{Mehdipour2017}
{Mehdipour}, M., {Kaastra}, J.~S., {Kriss}, G.~A., {et~al.} 2017, \aap, 607,
  A28

\bibitem[{{Mehdipour} {et~al.}(2021){Mehdipour}, {Kriss}, {Kaastra}, {Wang},
  {Mao}, {Costantini}, {Arav}, {Behar}, {Bianchi}, {Branduardi-Raymont},
  {Brotherton}, {Cappi}, {De Marco}, {Di Gesu}, {Ebrero}, {Grafton-Waters},
  {Kaspi}, {Matt}, {Paltani}, {Petrucci}, {Pinto}, {Ponti}, {Ursini}, \&
  {Walton}}]{Mehdipour2021}
{Mehdipour}, M., {Kriss}, G.~A., {Kaastra}, J.~S., {et~al.} 2021, \aap, 652,
  A150

\bibitem[{{Mineshige} {et~al.}(2000){Mineshige}, {Kawaguchi}, {Takeuchi}, \&
  {Hayashida}}]{Mineshige2000}
{Mineshige}, S., {Kawaguchi}, T., {Takeuchi}, M., \& {Hayashida}, K. 2000,
  \pasj, 52, 499

\bibitem[{{Miniutti} {et~al.}(2012){Miniutti}, {Brandt}, {Schneider},
  {et~al.}}]{Miniutti2012}
{Miniutti}, G., {Brandt}, W.~N., {Schneider}, D.~P., {et~al.} 2012, \mnras,
  425, 1718

\bibitem[{{Murray} {et~al.}(1995){Murray}, {Chiang}, {Grossman}, \&
  {Voit}}]{Murray1995}
{Murray}, N., {Chiang}, J., {Grossman}, S.~A., \& {Voit}, G.~M. 1995, \apj,
  451, 498

\bibitem[{{Nandra} {et~al.}(2013){Nandra}, {Barret}, {Barcons},
  {et~al.}}]{Nandra2013}
{Nandra}, K., {Barret}, D., {Barcons}, X., {et~al.} 2013, arXiv e-prints,
  arXiv:1306.2307.
\newblock \doarXiv{1306.2307}

\bibitem[{{Nardini} {et~al.}(2019){Nardini}, {Lusso}, {Risaliti}, {Bisogni},
  {Civano}, {Elvis}, {Fabbiano}, {Gilli}, {Marconi}, {Salvestrini}, \&
  {Vignali}}]{Nardini2019}
{Nardini}, E., {Lusso}, E., {Risaliti}, G., {et~al.} 2019, \aap, 632, A109

\bibitem[{{Netzer}(1993)}]{Netzer1993}
{Netzer}, H. 1993, \apj, 411, 594

\bibitem[{{Netzer}(2015)}]{Netzer2015}
---. 2015, \araa, 53, 365

\bibitem[{{Netzer} \& {Marziani}(2010)}]{Netzer2010}
{Netzer}, H., \& {Marziani}, P. 2010, \apj, 724, 318

\bibitem[{{Neugebauer} {et~al.}(1987){Neugebauer}, {Green}, {Matthews},
  {et~al.}}]{Neugebauer1987}
{Neugebauer}, G., {Green}, R.~F., {Matthews}, K., {et~al.} 1987, \apjs, 63, 615

\bibitem[{{Ni} {et~al.}(2018){Ni}, {Brandt}, {Luo}, {et~al.}}]{Ni2018}
{Ni}, Q., {Brandt}, W.~N., {Luo}, B., {et~al.} 2018, \mnras, 480, 5184

\bibitem[{{Ni} {et~al.}(2022){Ni}, {Brandt}, {Luo}, {et~al.}}]{Ni2022}
---. 2022, \mnras, 511, 5251

\bibitem[{{Ni} {et~al.}(2020){Ni}, {Brandt}, {Yi}, {et~al.}}]{Ni2020}
{Ni}, Q., {Brandt}, W.~N., {Yi}, W., {et~al.} 2020, \apjl, 889, L37

\bibitem[{{O'Donnell}(1994)}]{O'Donnell1994}
{O'Donnell}, J.~E. 1994, \apj, 422, 158

\bibitem[{{Ohsuga} \& {Mineshige}(2011)}]{Ohsuga2011}
{Ohsuga}, K., \& {Mineshige}, S. 2011, \apj, 736, 2

\bibitem[{{Panagiotou} \& {Walter}(2019)}]{Panagiotou2019}
{Panagiotou}, C., \& {Walter}, R. 2019, \aap, 626, A40

\bibitem[{{Park} {et~al.}(2006){Park}, {Kashyap}, {Siemiginowska}, {van Dyk},
  {Zezas}, {Heinke}, \& {Wargelin}}]{Park2006}
{Park}, T., {Kashyap}, V.~L., {Siemiginowska}, A., {et~al.} 2006, \apj, 652,
  610

\bibitem[{{Planck Collaboration} {et~al.}(2020){Planck Collaboration},
  {Aghanim}, {Akrami}, {Ashdown}, {Aumont}, {Baccigalupi}, {Ballardini},
  {Banday}, {Barreiro}, {Bartolo}, {Basak}, {Battye}, {Benabed}, {Bernard},
  {Bersanelli}, {Bielewicz}, {Bock}, {Bond}, {Borrill}, {Bouchet}, {Boulanger},
  {Bucher}, {Burigana}, {Butler}, {Calabrese}, {Cardoso}, {Carron},
  {Challinor}, {Chiang}, {Chluba}, {Colombo}, {Combet}, {Contreras}, {Crill},
  {Cuttaia}, {de Bernardis}, {de Zotti}, {Delabrouille}, {Delouis}, {Di
  Valentino}, {Diego}, {Dor{\'e}}, {Douspis}, {Ducout}, {Dupac}, {Dusini},
  {Efstathiou}, {Elsner}, {En{\ss}lin}, {Eriksen}, {Fantaye}, {Farhang},
  {Fergusson}, {Fernandez-Cobos}, {Finelli}, {Forastieri}, {Frailis},
  {Fraisse}, {Franceschi}, {Frolov}, {Galeotta}, {Galli}, {Ganga},
  {G{\'e}nova-Santos}, {Gerbino}, {Ghosh}, {Gonz{\'a}lez-Nuevo}, {G{\'o}rski},
  {Gratton}, {Gruppuso}, {Gudmundsson}, {Hamann}, {Handley}, {Hansen},
  {Herranz}, {Hildebrandt}, {Hivon}, {Huang}, {Jaffe}, {Jones}, {Karakci},
  {Keih{\"a}nen}, {Keskitalo}, {Kiiveri}, {Kim}, {Kisner}, {Knox},
  {Krachmalnicoff}, {Kunz}, {Kurki-Suonio}, {Lagache}, {Lamarre}, {Lasenby},
  {Lattanzi}, {Lawrence}, {Le Jeune}, {Lemos}, {Lesgourgues}, {Levrier},
  {Lewis}, {Liguori}, {Lilje}, {Lilley}, {Lindholm}, {L{\'o}pez-Caniego},
  {Lubin}, {Ma}, {Mac{\'\i}as-P{\'e}rez}, {Maggio}, {Maino}, {Mandolesi},
  {Mangilli}, {Marcos-Caballero}, {Maris}, {Martin}, {Martinelli},
  {Mart{\'\i}nez-Gonz{\'a}lez}, {Matarrese}, {Mauri}, {McEwen}, {Meinhold},
  {Melchiorri}, {Mennella}, {Migliaccio}, {Millea}, {Mitra},
  {Miville-Desch{\^e}nes}, {Molinari}, {Montier}, {Morgante}, {Moss}, {Natoli},
  {N{\o}rgaard-Nielsen}, {Pagano}, {Paoletti}, {Partridge}, {Patanchon},
  {Peiris}, {Perrotta}, {Pettorino}, {Piacentini}, {Polastri}, {Polenta},
  {Puget}, {Rachen}, {Reinecke}, {Remazeilles}, {Renzi}, {Rocha}, {Rosset},
  {Roudier}, {Rubi{\~n}o-Mart{\'\i}n}, {Ruiz-Granados}, {Salvati}, {Sandri},
  {Savelainen}, {Scott}, {Shellard}, {Sirignano}, {Sirri}, {Spencer},
  {Sunyaev}, {Suur-Uski}, {Tauber}, {Tavagnacco}, {Tenti}, {Toffolatti},
  {Tomasi}, {Trombetti}, {Valenziano}, {Valiviita}, {Van Tent}, {Vibert},
  {Vielva}, {Villa}, {Vittorio}, {Wandelt}, {Wehus}, {White}, {White},
  {Zacchei}, \& {Zonca}}]{Planck2020}
{Planck Collaboration}, {Aghanim}, N., {Akrami}, Y., {et~al.} 2020, \aap, 641,
  A6

\bibitem[{{Proga} {et~al.}(2000){Proga}, {Stone}, \& {Kallman}}]{Proga2000}
{Proga}, D., {Stone}, J.~M., \& {Kallman}, T.~R. 2000, \apj, 543, 686

\bibitem[{{Pu} {et~al.}(2020){Pu}, {Luo}, {Brandt}, {Timlin}, {Liu}, {Ni}, \&
  {Wu}}]{Pu2020}
{Pu}, X., {Luo}, B., {Brandt}, W.~N., {et~al.} 2020, \apj, 900, 141

\bibitem[{{Reeves} {et~al.}(1997){Reeves}, {Turner}, {Ohashi}, \&
  {Kii}}]{Reeves1997}
{Reeves}, J.~N., {Turner}, M.~J.~L., {Ohashi}, T., \& {Kii}, T. 1997, \mnras,
  292, 468

\bibitem[{{Ricci} {et~al.}(2017){Ricci}, {Trakhtenbrot}, {Koss},
  {et~al.}}]{Ricci2017}
{Ricci}, C., {Trakhtenbrot}, B., {Koss}, M.~J., {et~al.} 2017, \apjs, 233, 17

\bibitem[{{Ricci} {et~al.}(2011){Ricci}, {Walter}, {Courvoisier}, \&
  {Paltani}}]{Ricci2011}
{Ricci}, C., {Walter}, R., {Courvoisier}, T.~J.~L., \& {Paltani}, S. 2011,
  \aap, 532, A102

\bibitem[{{Ricci} {et~al.}(2021){Ricci}, {Loewenstein}, {Kara}, {Remillard},
  {Trakhtenbrot}, {Arcavi}, {Gendreau}, {Arzoumanian}, {Fabian}, {Li}, {Ho},
  {MacLeod}, {Cackett}, {Altamirano}, {Gandhi}, {Kosec}, {Pasham}, {Steiner},
  \& {Chan}}]{Ricci2021}
{Ricci}, C., {Loewenstein}, M., {Kara}, E., {et~al.} 2021, \apjs, 255, 7

\bibitem[{{Ross} \& {Fabian}(2005)}]{Ross2005}
{Ross}, R.~R., \& {Fabian}, A.~C. 2005, \mnras, 358, 211

\bibitem[{{Rovilos} {et~al.}(2014){Rovilos}, {Georgantopoulos}, {Akylas},
  {et~al.}}]{Rovilos2014}
{Rovilos}, E., {Georgantopoulos}, I., {Akylas}, A., {et~al.} 2014, \mnras, 438,
  494

\bibitem[{{Sabra} \& {Hamann}(2001)}]{Sabra2001}
{Sabra}, B.~M., \& {Hamann}, F. 2001, \apj, 563, 555

\bibitem[{{Saez} {et~al.}(2012){Saez}, {Brandt}, {Gallagher}, {Bauer}, \&
  {Garmire}}]{Saez2012}
{Saez}, C., {Brandt}, W.~N., {Gallagher}, S.~C., {Bauer}, F.~E., \& {Garmire},
  G.~P. 2012, \apj, 759, 42

\bibitem[{{Schartel} {et~al.}(2005){Schartel}, {Rodr{\'\i}guez-Pascual},
  {Santos-Lle{\'o}}, {Clavel}, {Guainazzi}, {Jim{\'e}nez-Bail{\'o}n}, \&
  {Piconcelli}}]{Schartel2005}
{Schartel}, N., {Rodr{\'\i}guez-Pascual}, P.~M., {Santos-Lle{\'o}}, M.,
  {et~al.} 2005, \aap, 433, 455

\bibitem[{{Scott} {et~al.}(2011){Scott}, {Stewart}, {Mateos}, {Alexander},
  {Hutton}, \& {Ward}}]{Scott2011}
{Scott}, A.~E., {Stewart}, G.~C., {Mateos}, S., {et~al.} 2011, \mnras, 417, 992

\bibitem[{{Shemmer} {et~al.}(2008){Shemmer}, {Brandt}, {Netzer}, {Maiolino}, \&
  {Kaspi}}]{Shemmer2008}
{Shemmer}, O., {Brandt}, W.~N., {Netzer}, H., {Maiolino}, R., \& {Kaspi}, S.
  2008, \apj, 682, 81

\bibitem[{{Shen} \& {Ho}(2014)}]{Shen2014}
{Shen}, Y., \& {Ho}, L.~C. 2014, \nat, 513, 210

\bibitem[{{Shen} {et~al.}(2011){Shen}, {Richards}, {Strauss}, {Hall},
  {Schneider}, {Snedden}, {Bizyaev}, {Brewington}, {Malanushenko},
  {Malanushenko}, {Oravetz}, {Pan}, \& {Simmons}}]{shen2011}
{Shen}, Y., {Richards}, G.~T., {Strauss}, M.~A., {et~al.} 2011, \apjs, 194, 45

\bibitem[{{Skrutskie} {et~al.}(2006){Skrutskie}, {Cutri}, {Stiening},
  {Weinberg}, {Schneider}, {Carpenter}, {Beichman}, {Capps}, {Chester},
  {Elias}, {Huchra}, {Liebert}, {Lonsdale}, {Monet}, {Price}, {Seitzer},
  {Jarrett}, {Kirkpatrick}, {Gizis}, {Howard}, {Evans}, {Fowler}, {Fullmer},
  {Hurt}, {Light}, {Kopan}, {Marsh}, {McCallon}, {Tam}, {Van Dyk}, \&
  {Wheelock}}]{Skrutskie2006}
{Skrutskie}, M.~F., {Cutri}, R.~M., {Stiening}, R., {et~al.} 2006, \aj, 131,
  1163

\bibitem[{{Steffen} {et~al.}(2006){Steffen}, {Strateva}, {Brandt}, {Alexander},
  {Koekemoer}, {Lehmer}, {Schneider}, \& {Vignali}}]{steffen2006}
{Steffen}, A.~T., {Strateva}, I., {Brandt}, W.~N., {et~al.} 2006, \aj, 131,
  2826

\bibitem[{{Stern}(2015)}]{Stern2015}
{Stern}, D. 2015, \apj, 807, 129

\bibitem[{{Strateva} {et~al.}(2005){Strateva}, {Brandt}, {Schneider}, {Vanden
  Berk}, \& {Vignali}}]{Strateva2005}
{Strateva}, I.~V., {Brandt}, W.~N., {Schneider}, D.~P., {Vanden Berk}, D.~G.,
  \& {Vignali}, C. 2005, \aj, 130, 387

\bibitem[{{Sulentic} {et~al.}(2000){Sulentic}, {Zwitter}, {Marziani}, \&
  {Dultzin-Hacyan}}]{Sulentic2000}
{Sulentic}, J.~W., {Zwitter}, T., {Marziani}, P., \& {Dultzin-Hacyan}, D. 2000,
  \apjl, 536, L5

\bibitem[{{Takeuchi} {et~al.}(2014){Takeuchi}, {Ohsuga}, \&
  {Mineshige}}]{Takeuchi2014}
{Takeuchi}, S., {Ohsuga}, K., \& {Mineshige}, S. 2014, \pasj, 66, 48

\bibitem[{{Tananbaum} {et~al.}(1979){Tananbaum}, {Avni}, {Branduardi}, {Elvis},
  {Fabbiano}, {Feigelson}, {Giacconi}, {Henry}, {Pye}, {Soltan}, \&
  {Zamorani}}]{Tananbaum1979}
{Tananbaum}, H., {Avni}, Y., {Branduardi}, G., {et~al.} 1979, \apjl, 234, L9

\bibitem[{{Teng} {et~al.}(2014){Teng}, {Brandt}, {Harrison}, {Luo},
  {Alexander}, {Bauer}, {Boggs}, {Christensen}, {Comastri}, {Craig}, {Fabian},
  {Farrah}, {Fiore}, {Gandhi}, {Grefenstette}, {Hailey}, {Hickox}, {Madsen},
  {Ptak}, {Rigby}, {Risaliti}, {Saez}, {Stern}, {Veilleux}, {Walton}, {Wik}, \&
  {Zhang}}]{Teng2014}
{Teng}, S.~H., {Brandt}, W.~N., {Harrison}, F.~A., {et~al.} 2014, \apj, 785, 19

\bibitem[{{Trakhtenbrot} {et~al.}(2019){Trakhtenbrot}, {Arcavi}, {MacLeod},
  {Ricci}, {Kara}, {Graham}, {Stern}, {Harrison}, {Burke}, {Hiramatsu},
  {Hosseinzadeh}, {Howell}, {Smartt}, {Rest}, {Prieto}, {Shappee}, {Holoien},
  {Bersier}, {Filippenko}, {Brink}, {Zheng}, {Li}, {Remillard}, \&
  {Loewenstein}}]{Trakhtenbrot2019}
{Trakhtenbrot}, B., {Arcavi}, I., {MacLeod}, C.~L., {et~al.} 2019, \apj, 883,
  94

\bibitem[{{Turner} {et~al.}(1997){Turner}, {George}, {Nandra}, \&
  {Mushotzky}}]{Turner1997}
{Turner}, T.~J., {George}, I.~M., {Nandra}, K., \& {Mushotzky}, R.~F. 1997,
  \apjs, 113, 23

\bibitem[{{Turner} \& {Miller}(2009)}]{Turner2009}
{Turner}, T.~J., \& {Miller}, L. 2009, \aapr, 17, 47

\bibitem[{{Ueda} {et~al.}(2007){Ueda}, {Eguchi}, {Terashima}, {Mushotzky},
  {Tueller}, {Markwardt}, \& {Gehrels}}]{Ueda2007}
{Ueda}, Y., {Eguchi}, S., {Terashima}, Y., {et~al.} 2007, Progress of
  Theoretical Physics Supplement, 169, 295

\bibitem[{{Vanden Berk} {et~al.}(2001){Vanden Berk}, {Richards}, {Bauer},
  {et~al.}}]{Vandenberk2001}
{Vanden Berk}, D.~E., {Richards}, G.~T., {Bauer}, A., {et~al.} 2001, \aj, 122,
  549

\bibitem[{{Veilleux} {et~al.}(2009){Veilleux}, {Rupke}, {Kim},
  {et~al.}}]{Veilleux2009}
{Veilleux}, S., {Rupke}, D.~S.~N., {Kim}, D.~C., {et~al.} 2009, \apjs, 182, 628

\bibitem[{{Wang} \& {Netzer}(2003)}]{Wang2003}
{Wang}, J.-M., \& {Netzer}, H. 2003, \aap, 398, 927

\bibitem[{{Wills} {et~al.}(2000){Wills}, {Shang}, \& {Yuan}}]{Wills2000}
{Wills}, B.~J., {Shang}, Z., \& {Yuan}, J.~M. 2000, \nar, 44, 511

\bibitem[{{Winter} {et~al.}(2009){Winter}, {Mushotzky}, {Reynolds}, \&
  {Tueller}}]{Winter2009}
{Winter}, L.~M., {Mushotzky}, R.~F., {Reynolds}, C.~S., \& {Tueller}, J. 2009,
  \apj, 690, 1322

\bibitem[{{Wright} {et~al.}(2010){Wright}, {Eisenhardt}, {Mainzer}, {Ressler},
  {Cutri}, {Jarrett}, {Kirkpatrick}, {Padgett}, {McMillan}, {Skrutskie},
  {Stanford}, {Cohen}, {Walker}, {Mather}, {Leisawitz}, {Gautier}, {McLean},
  {Benford}, {Lonsdale}, {Blain}, {Mendez}, {Irace}, {Duval}, {Liu}, {Royer},
  {Heinrichsen}, {Howard}, {Shannon}, {Kendall}, {Walsh}, {Larsen}, {Cardon},
  {Schick}, {Schwalm}, {Abid}, {Fabinsky}, {Naes}, \& {Tsai}}]{Wright2010}
{Wright}, E.~L., {Eisenhardt}, P. R.~M., {Mainzer}, A.~K., {et~al.} 2010, \aj,
  140, 1868

\bibitem[{{Yamada} {et~al.}(2020){Yamada}, {Ueda}, {Tanimoto}, {Oda},
  {Imanishi}, {Toba}, \& {Ricci}}]{Yamada2020}
{Yamada}, S., {Ueda}, Y., {Tanimoto}, A., {et~al.} 2020, \apj, 897, 107

\bibitem[{{Yi} {et~al.}(2022){Yi}, {Brandt}, {Ni}, {Ho}, {Luo}, {Yan},
  {Schneider}, {Paul}, {Plotkin}, {Yang}, {Wang}, {He}, {Chen}, {Wu}, \&
  {Bai}}]{Yi2022}
{Yi}, W., {Brandt}, W.~N., {Ni}, Q., {et~al.} 2022, \apj, 930, 5

\bibitem[{{York} {et~al.}(2000){York}, {Adelman}, {Anderson}, {Anderson},
  {Annis}, {Bahcall}, {Bakken}, {Barkhouser}, {Bastian}, {Berman}, {Boroski},
  {Bracker}, {Briegel}, {Briggs}, {Brinkmann}, {Brunner}, {Burles}, {Carey},
  {Carr}, {Castander}, {Chen}, {Colestock}, {Connolly}, {Crocker}, {Csabai},
  {Czarapata}, {Davis}, {Doi}, {Dombeck}, {Eisenstein}, {Ellman}, {Elms},
  {Evans}, {Fan}, {Federwitz}, {Fiscelli}, {Friedman}, {Frieman}, {Fukugita},
  {Gillespie}, {Gunn}, {Gurbani}, {de Haas}, {Haldeman}, {Harris}, {Hayes},
  {Heckman}, {Hennessy}, {Hindsley}, {Holm}, {Holmgren}, {Huang}, {Hull},
  {Husby}, {Ichikawa}, {Ichikawa}, {Ivezi{\'c}}, {Kent}, {Kim}, {Kinney},
  {Klaene}, {Kleinman}, {Kleinman}, {Knapp}, {Korienek}, {Kron}, {Kunszt},
  {Lamb}, {Lee}, {Leger}, {Limmongkol}, {Lindenmeyer}, {Long}, {Loomis},
  {Loveday}, {Lucinio}, {Lupton}, {MacKinnon}, {Mannery}, {Mantsch}, {Margon},
  {McGehee}, {McKay}, {Meiksin}, {Merelli}, {Monet}, {Munn}, {Narayanan},
  {Nash}, {Neilsen}, {Neswold}, {Newberg}, {Nichol}, {Nicinski}, {Nonino},
  {Okada}, {Okamura}, {Ostriker}, {Owen}, {Pauls}, {Peoples}, {Peterson},
  {Petravick}, {Pier}, {Pope}, {Pordes}, {Prosapio}, {Rechenmacher}, {Quinn},
  {Richards}, {Richmond}, {Rivetta}, {Rockosi}, {Ruthmansdorfer}, {Sandford},
  {Schlegel}, {Schneider}, {Sekiguchi}, {Sergey}, {Shimasaku}, {Siegmund},
  {Smee}, {Smith}, {Snedden}, {Stone}, {Stoughton}, {Strauss}, {Stubbs},
  {SubbaRao}, {Szalay}, {Szapudi}, {Szokoly}, {Thakar}, {Tremonti}, {Tucker},
  {Uomoto}, {Vanden Berk}, {Vogeley}, {Waddell}, {Wang}, {Watanabe},
  {Weinberg}, {Yanny}, {Yasuda}, \& {SDSS Collaboration}}]{York2000}
{York}, D.~G., {Adelman}, J., {Anderson}, John~E., J., {et~al.} 2000, \aj, 120,
  1579

\bibitem[{{Yuan} \& {Wills}(2003)}]{Yuan2003}
{Yuan}, M.~J., \& {Wills}, B.~J. 2003, \apjl, 593, L11

\end{thebibliography}

\appendix
\section{XMM-Newton Observation of LBQS~$1442-0011$}
LBQS~$1442-0011$ is a BAL quasar at $z = 2.226$ with a $B$--band magnitude of 18.2 
\citep{Gallagher2006}. Its H$\beta$-based single-epoch 
virial SMBH mass is $\approx8\times10^{9} M_{\Sun}$, and its estimated 
Eddington ratio is $\approx0.17$ \citep{Yuan2003}. 
Its previous \chandra\ observations have a co-added depth of 15.9~ks. Through systematic
analyses of the \chandra\ observations of the \citet{Gallagher2006} LBQS BAL quasar sample,
\citet{Liu2018} identified LBQS~$1442-0011$ as one of the two good candidates for 
intrinsically X-ray weak quasars based on its significant hard 
X-ray weakness (by a factor of $f_{\rm w}=12^{+12}_{-8}$) 
and its nominal hard X-ray spectral shape ($\Gamma_{\rm eff} = 1.9^{+0.9}_{-0.8}$). The other 
candidate is LBQS~$1203+1530$.

Due to the large uncertainties of the $f_{\rm w}$ and $\Gamma_{\rm eff}$ values from the 
\chandra\ results, we proposed for 
deeper \xmm\ observations of the two candidates, aiming to improve the parameter
constraints. The targets were accepted at priority C, and LBQS~1442--0011 was observed on
2021 February 6 with a nominal exposure time of 87~ks. Unfortunately, 
the observation was affected significantly by background flares, 
and the cleaned exposure time is only 34~ks. Thus the sensitivity of the new \xmm\ observation
is only comparable to the previous co-added \chandra\ exposure. 
We processed the \xmm\ data following the procedure described in Section~2.3. 
We chose a smaller source extraction region with 
a radius of $25\arcsec$ in order to increase the signal-to-noise ratio 
of this faint source. We also limited the upper-energy bound to 8~keV. 
The resulting net source counts 
are $48^{+18}_{-17}$ in the 0.3--2~keV band (rest-frame 1.0--6.5 keV) 
and $33^{+15}_{-14}$ in the 2--8~keV band (rest-frame 6.5--26 keV).
The $\Gamma_{\rm eff,0.3-8}$ value inferred from the 
band ratio is $1.0^{+0.7}_{-0.6}$. 
We also fit the spectrum with a power-law model modified by Galactic absorption 
(\texttt{phabs}\,*\,\texttt{zpow}), and the best-fit $\Gamma$ value ($0.9\pm0.4$) 
is consistent with the photometric result. The derived factor of hard X-ray weakness is 
$f_{\rm w}=4\pm2$.

Compared to the previous \chandra\ constraints, the \xmm\ results suggest that the hard X-ray
spectrum
became flatter, and the observed hard X-ray emission became brighter ($f_{\rm w}$ dropped).
Therefore, we suggest that the X-ray weakness of 
LBQS~$1442-0011$ is also caused by variable partial-covering absorption, similar to 
the three quasars studied here (see Section~4.3 for discussion).

\end{document}